\documentclass[11pt]{article}
\usepackage{booktabs}
\usepackage{float}
\usepackage{amssymb,amsmath,amsfonts, mathtools}

\usepackage{enumerate}
\usepackage{dsfont}
\usepackage[sort,compress]{cite}
\usepackage{verbatim}
\usepackage{amsfonts,euscript,amssymb,stmaryrd,braket}
\usepackage{graphics,tikz}
\usetikzlibrary{arrows,decorations.markings,patterns}
\usepackage{caption}
\usepackage{subcaption}
\usepackage{slashed}
\definecolor{hyperref}{RGB}{026,028,185}
\usepackage[bookmarks=true,colorlinks=true,linkcolor=hyperref,citecolor=hyperref,urlcolor=hyperref,bookmarksnumbered]{hyperref}
 \usepackage{booktabs}
 \usepackage{multirow}
\usepackage{tabularx}
\usepackage{longtable}
\usepackage{bbold,bbding}
\usepackage{amsthm} 
\usepackage{graphicx}

 \def\clock{{\count0=\time
           \divide\count0 60
           \ifnum\count0<10 0\fi\the\count0
           \multiply\count0 -60 \advance\count0 \time
           :\ifnum\count0<10 0\fi \the\count0
         }}
\newcommand{\timestamp}{{\small\vbox{\hbox{\tt\jobname.tex}
\hbox{\the\day/\the\month/\the\year, \clock}}}}

\def\EV#1{{\color [rgb]{0.8,0.2,0} [EV: #1]}}

\newcommand{\rf}[1]{(\ref{#1})}

\newcommand{\ba}{\begin{eqnarray}}
\newcommand{\ea}{\end{eqnarray}}

\newcommand{\tr}{\textrm{tr}}

\newcommand{\Det}{\textrm{Det}}

\newcommand{\arctanh}{\textrm{arctanh}}

\newcommand{\be}{\begin{equation}}
\newcommand{\ee}{\end{equation}}

\makeatletter
\let\old@startsection=\@startsection
\let\oldl@section=\l@section
\renewcommand{\@startsection}[6]{\old@startsection{#1}{#2}{#3}{#4}{#5}{#6\mathversion{bold}}}
\renewcommand{\l@section}[2]{\oldl@section{\mathversion{bold}#1}{#2}}
\makeatother

\numberwithin{equation}{section}

\usepackage{color}

\def\la{\label}
\def\no{\nonumber}

\begin{document}
\renewcommand{\thefootnote}{\arabic{footnote}}
 
\date{\currenttime}
\date {}
\vspace{.2cm}

\overfullrule=0pt
\parskip=2pt
\parindent=12pt
\headheight=0in \headsep=0in \topmargin=0in \oddsidemargin=0in

\vspace{ 1cm} \thispagestyle{empty} \vspace{-1cm}
\begin{flushright} 
\footnotesize
HU-EP-17/02
\\
Imperial-TP-AAT-2017-03
\end{flushright}%

\begin{center}
\vspace{1.2cm}
{\Large\bf \mathversion{bold}
Perturbative computation of string one-loop corrections 
 \vspace{0.2cm}
to
 Wilson loop   minimal  surfaces in $AdS_5 \times S^5$ 
}

 \vspace{0.8cm} {
 V.~Forini$^{a,}$\footnote{{\tt valentina.forini@\,physik.hu-berlin.de}}, A.A.~Tseytlin$^{b,}$\footnote{Also at Lebedev Institute, Moscow.  {\tt  tseytlin@imperial.ac.uk}}  and    E.~Vescovi$^{a,c,}$\footnote{{\tt vescovi@\,if.usp.br}}}
 \vskip  0.5cm

\small
{\em
$^{a}$Institut f\"ur Physik, Humboldt-Universit\"at zu Berlin, IRIS Adlershof, \\Zum Gro\ss en Windkanal 6, 12489 Berlin, Germany  
\vskip 0.05cm
$^{b}$Theoretical Physics Group, Blackett Laboratory, Imperial College, London, SW7 2AZ, U.K.
\vskip 0.05cm
$^{c}$Institute of Physics, University of S\~{a}o Paulo, Rua do Mat\~{a}o 1371, 05508-090 S\~{a}o Paulo, Brazil
}
\normalsize

\end{center}

\vspace{0.3cm}
\begin{abstract} 
We revisit the computation of the 1-loop  string correction to the ``latitude"   minimal surface  in $AdS_5 \times S^5$   representing 
1/4 BPS Wilson loop in planar $\cal N$=4   SYM theory  previously addressed in arXiv:1512.00841  and arXiv:1601.04708. 
We resolve the problem  of matching with the subleading term in the strong coupling  expansion of the 
exact gauge theory result (derived previously from localization)  using   a different method to compute  determinants 
of 2d   string fluctuation operators. We apply  perturbation theory  in a small parameter (angle of the latitude) 
corresponding to  an expansion  near the  $AdS_2$  minimal surface  representing 1/2 BPS  circular Wilson loop. This  
allows us  to compute   the corrections  to the heat kernels  and  zeta-functions  of the operators   in terms of the 
known  heat kernels on $AdS_2$. We apply   the same method  also to two other examples of Wilson 
loop surfaces:   generalized   cusp and $k$-wound circle. 

\end{abstract}

\newpage

\tableofcontents
  
  \setcounter{footnote}{0}\def \ci{\cite} \def \foot{\footnote}\def \a {\alpha} 
  \def \W {\mathcal{W}} \def \l {\lambda} \def \te {\textstyle} \def \ov {\over}
  \def \ha {\te {1\ov 2}}
\def \te {\textstyle}
\def \ve {\Lambda}
\def \s {\sigma} \def \ov {\over} 
\def \ci {\cite} \def \td {\tilde} 
  \def \a {\alpha}
\def \iffa {\iffalse} 
\def \foot {\footnote}
\def \ed {\end{document}}

\section{Introduction}

The expectation   value   of a Wilson loop (WL) operator  in  planar $\mathcal{N}=4$  super Yang-Mills   theory is conjected  to be 
 given, at strong coupling,    by the  $AdS_5 \times S^5  $ superstring  path integral   with  appropriate   boundary conditions
 \ci{Maldacena:1998im,Rey:1998ik,Berenstein:1998ij}. 
 The computation of the leading  strong-coupling correction to the classical area term  
 given by the  logarithm  of the 1-loop   string partition function was addressed in \ci{Kallosh:1998ji,Forste:1999qn,Drukker:2000ep}
 and, in general, is   technically challenging. 
 
Simplest   examples  correspond to supersymmetric Wilson  loops, e.g., 
  1/2 BPS circular loop ~\cite{Drukker:2000ep, Kruczenski:2008zk,Buchbinder:2014nia},  
 1/4  BPS family of  ``latitudes''~\cite{Drukker:2005cu,Drukker:2006ga,Drukker:2007qr,Forini:2015mca,Forini:2015bgo,Faraggi:2016ekd}, 
  the $k$-wound  circle case (dual to WL  in  $k$-fundamental representation) ~\cite{Kruczenski:2008zk,Bergamin:2015vxa},  etc. 
Even   in the   circular WL case   the first string correction appears to
 disagree with   the subleading term in the strong coupling expansion of the gauge-theory result~\cite{Erickson:2000af, Semenoff:2002kk, Drukker:2000rr,Drukker:2005kx,Pestun:2007rz,Pestun:2009nn,Zarembo:2016bbk}. 
 
To avoid the   subtle issue   of the overall normalization of the  string path integral  one may  consider the computation of the {\it  ratio}
of  partition functions for minimal surfaces of the same  (disc)  topology. Then
 the universal UV divergences   and  possible 
 string tension  factors associated with   the Killing  vector volume  \ci{Drukker:2000ep,Drukker:2000rr}  
 that are independent of local  world-sheet geometry should   cancel out  and the result should be  a well-defined function of the 
 non-trivial  WL (i.e. world-surface) parameters.  
This   strategy was  followed   in ~\cite{Forini:2015bgo}, where the
 one-loop determinants for   fluctuations about the classical string solutions corresponding to a generic  1/4  BPS  
  ``latitude'' WL  of~\cite{Drukker:2005cu,Drukker:2006ga,Drukker:2007qr} 
 were evaluated with the Gel'fand-Yaglom  (GY) method. 
  The same  result   for the  string partition function 
  was obtained in~\cite{Faraggi:2016ekd}, with a slightly different application 
  of the  GY  method.\footnote{In~\cite{Faraggi:2016ekd},  
the fermionic contribution was  found   starting  with 
 the Dirac-like first-order operator rather than  its square, as in~\cite{Forini:2015bgo}.  
 Using  a  particular  organization  of the determinant ratios,   ref. \cite{Faraggi:2016ekd}   computed   the analytic 
 expression for the resulting string 1-loop correction
   (while the analysis in~\cite{Forini:2015bgo} was partially numerical).
    Ref. \cite{Faraggi:2016ekd} presented 
    also a  detailed study  of the supermultiplet structure of the fluctuations.
   } 
Still, the  resulting string  prediction   was found to  be 
 in  disagreement with the exact gauge theory result   obtained  by the  localization method \ci{Pestun:2007rz,Pestun:2009nn}.

In this paper we will reconsider the  computation in \ci{Forini:2015bgo,Faraggi:2016ekd}  using a different approach  to evaluation of  the 
fluctuation determinants. We shall use  the perturbation theory in a small parameter $\a$, 
such that for $\a=0$   the  world-surface becomes the same as 
 the circular WL  surface, i.e. is   equivalent to the Euclidean $AdS_2$. Then the  leading correction in $\a$  can be  found  by 
the perturbative expansion of the  heat  kernels (see,  e.g.,  \cite{GamboaSaravi:1983ofo, Mukhanov})    using  that  for  $\a=0$, i.e. in the $AdS_2$ case, the 
 heat kernels  for the bosonic and fermionic operators   are known explicitly  ~\cite{Camporesi:1990wm,Camporesi:1994ga,Camporesi:1992tm,Camporesi:1995fb}.
This  will  allow us to  find the  leading-order   correction to the  string partition function 
for  the near-$AdS_2$ geometry  corresponding to  the  latitude in $S^2\subset  S^5$ parametrized by a small angle $\theta_0$.
Since for  $\theta_0=0$   it reduces to the $AdS_2$ (circular WL)    geometry,  here 
  the small expansion parameter   may be  chosen  as $\a= \theta^2_0$. 

 Remarkably, we  will be able  to reproduce the  first   non-trivial  term   in  the  small-$\theta_0$ expansion of the 
 exact gauge-theory   result~\ci{Pestun:2007rz,Pestun:2009nn} 
  for the latitude WL expectation value $Z=  \langle \W(\l,\theta_0)\rangle$  in the strong-coupling ($\lambda \gg 1$)   limit. 
 Explicitly,  the gauge-theory prediction  
   for the string ``effective action"    $\Gamma=-\log Z$ 
   is 
\be\label{1.1}
\!\!\!\!
\Gamma(\l,\theta_0)-\Gamma(\l,0)
=\sqrt{\lambda}\,(1-\cos\theta_0)+\te \frac{3}{2}\log\cos\theta_0+\mathcal{O} (\lambda^{-{1}/{2}} ) ~,
\ee
and we will   reproduce precisely   the  leading  small-$\theta_0$  term in the $O(\lambda^0)$  part of \rf{1.1}, i.e. 
 $ \frac{3}{2}\log\cos\theta_0= - { 3 \ov 4} \theta^2_0  + O(\theta^4_0)$,  
from the  one-loop  string-theory computation (see \rf{3.2},\rf{3.46}). 

A possible   reason why the  two  previous    attempts  in \ci{Forini:2015bgo} and \ci{Faraggi:2016ekd}  failed to find the   agreement with
the  gauge  theory result  may be 
related to  some subtleties  in their  application of the GY  method to computation of  functional determinants.\foot{This method   was  originally suggested in \cite{Gelfand:1959nq} and later improved in~\cite{Forman1987, Forman1992, McKane:1995vp, Kirsten:2003py, Kirsten:2004qv, Kirsten:2007ev}; for a review  see,  for example, ~\cite{Kirsten:2004qv,Dunne:2007rt}, or 
Appendix B of~\cite{Forini:2015bgo}.
}
 Compared to the heat-kernel approach, here 
  the spectral problem is  treated  (after Fourier-transforming in $\tau$) as effectively  a  one-dimensional operator problem; 
    one  also  uses a zeta-function-like regularization in $\sigma$  world-sheet  direction 
    and a cutoff regularization  of  the sum over the Fourier  modes in $\tau$-direction.
     This method  also requires considering ratios of determinants for differential operators with the same principal symbol, which in turns implies  a functional rescaling by a  conformal factor.\foot{One may quantify (see, e.g.,  Appendix A of~\cite{Drukker:2000ep})
      how such conformal rescaling of the operators affects the finite part of the regularized determinants. However, a simple check for the ratio of  two bosonic operators  in\cite{Forini:2015bgo, Faraggi:2016ekd} reveals that adding this contribution does not  explain  the discrepancy with the  result obtained here.} 
      Together with  a possible regularization ambiguity  in   the sum over modes mentioned above, 
      what may account for the   disagreement 
      is the fictitious boundary (a  cut at the origin of the disk)
        introduced in~\cite{Kruczenski:2008zk,Forini:2015bgo, Faraggi:2016ekd} to allow
        for  the calculation of determinants on a compact  interval (see also~\cite{Frolov:2004bh, Dekel:2013kwa}). 
          It would be interesting to perform an explicit comparison of  the two 
           computations  eliminating the need for  this regulator, which  does not appear  in the heat kernel approach.\footnote{A more general 
           application of the GY  method~\cite{Dunne:2006ct} suggests that in the case of a non-compact interval one may try to proceed 
            by selecting suitably ``well-behaved'' eigenfunctions of the auxiliary initial value problem.} 

Below we  will  also  test  our    perturbative approach   based on  constructing  heat kernels for 2d  fluctuation operators 
  in   an expansion in a small parameter  on two other examples.  
  The first  will be    the near-BPS limit of the generalized cusp of~\cite{Drukker:2011za},  corresponding to the  
 the strong coupling  expansion of the ``Bremsstrahlung function" of $\mathcal{N}=4$ SYM theory, 
 derived exactly  using  supersymmetric localization in  \cite{Correa:2012at}. 
In this case  the GY  method applied   to the computation 
of the string 1-loop   correction   reproduced ~\cite{Drukker:2011za}
 the   gauge-theory  result.\footnote{Here  the application of the GY method does not require 
   an  unphysical regulator    and thus the agreement could be expected.  
   The GY  procedure  is known  also   to  reproduce the predictions of integrability  on gauge-theory side 
    in other non-trivial fluctuation problems~\cite{Beccaria:2010ry, Forini:2010ek,Forini:2012bb, Forini:2014kza}.}
    Our perturbative computation will    also  be consistent with this matching. 
    
    Another example  will be   the 1-loop partition function  for the surface   ending on the  $k$-wound circle 
that should be representing  the $k$-fundamental  circular Wilson loop~\cite{Berenstein:1998ij,Drukker:1999zq}.
Here the  gauge theory result  is a  generalization of the $k=1$ circular WL case~\cite{Drukker:2005cu,Pestun:2007rz}, 
see \rf{3.107}.  
 The string   one-loop computation    was   previously discussed  
 in ~\cite{Kruczenski:2008zk}  (using the GY  method and again introducing an unphysical cutoff)  and in \cite{Bergamin:2015vxa} (using heat kernel  construction on a cone of $AdS_2$ with angular deficit $2\pi(1-k)$). Both approaches   failed
 to find  an   agreement with gauge theory. 
 We will use an  expansion about  the $k=1$ case, i.e.  set  the small  parameter  to be $\a=k-1$.
Our result \eqref{3.105}  for the  coefficient of the $O(k-1)$ term in the 1-loop correction 
 will differ from the  gauge theory one   just  by an extra $\gamma$-term (the Euler-Mascheroni constant). 
 We will suggest that this disagreement is due to a regularization ambiguity  related to 
 the fact that the expansion   near  the regular  $k=1$   (i.e. $AdS_2$) surface   
 appears to be problematic    due to  a conical  singularity appearing for $k \not=1$. 
 
 We will start  in Section~\ref{sec:perturbative_hk}
 with the description of the perturbative procedure   for computing the  heat kernel in a small-parameter expansion.
 In  Section~\ref{sec:applications}  we will apply this   method the   1-loop  string computations  of the leading 
 corrections to the three  WL surfaces mentioned above. 
  We  will collect useful formulae and details of the calculations in 
   Appendices \ref{app:pert_theory} and \ref{app:heat_kernel_zeta_function}. 
\section{Perturbative expansion of  heat kernel  and determinant of an elliptic operator} 
\label{sec:perturbative_hk}

To prepare for the  computation of  leading string 1-loop  corrections to Wilson loop expectation   values 
in expansion in some small parameter $\a$  here we shall present the general relations for the perturbative  expansion of the 
heat kernel  and determinant   of a  differential operator parametrized  by $\a$.

Let  $\mathcal{O}$   be a second order elliptic operator   defined on (sections of a  bundle over) a $d$-dimensional  Riemannian manifold $\mathcal{M}$  with metric $g_{ij}$. The standard   expression for  the  logarithm of its determinant defined using zeta-function regularization  is  (see, e.g., ~\cite{Vassilevich:2003xt,Fursaev}) 
\begin{eqnarray}\label{2.1}
&&\log\, \Det_{_{\mathcal{M}}}\mathcal{O} =-\zeta_{\mathcal{O}}^{'}\left(0\right)\,, \\
\label{2.2}
&& {\zeta}_{\mathcal{O}}\left(s\right)  =\frac{1}{\Gamma\left(s\right)}\int_{0}^{\infty}dt\,t^{s-1}\,{ K}_{\mathcal{O}}\left(t\right)\,,\qquad K_{\mathcal{O}}\left(t\right)  =\int d^d{x}\sqrt{g}~\tr\,  K_{\mathcal{O}}\left(x,x;t\right)~,
\\
\label{2.3}
&& \left(\partial_{t}+\mathcal{O}_{x}\right)K_{\mathcal{O}}(x,x{'};t)=0\ , \qquad\qquad~ K_{\mathcal{O}}(x,x{'};0)=\frac{1}{\sqrt{g}}\delta^{\left(d\right)}\big(x-x{'}\big)\, \mathbb{I}~.
\end{eqnarray}
Here  $\tr $  and  the unit operator $\mathbb{\mathbb{I}}$ 
      correspond to the internal indices  in the  (vector or spinor) bundle.

Suppose  the metric  $g_{ij}$ on $\mathcal{M}$  as well as $\mathcal{O}$   depend on some  parameter $\a$, 
such that for $\a=0$, corresponding to $\bar{\mathcal{M}}$ with metric $\bar{g}_{ij}$, the spectral problem can be solved exactly.  Then we can   compute $K_{\mathcal{O}}$    and $\Det_{_{\mathcal{M}}}\mathcal{O}$ in 
 perturbation theory in $\a$.   Namely, let us  set 
\begin{equation}\label{expansion}
\begin{split}
g_{ij}&=\bar{g}_{ij}+\alpha~\tilde{g}_{ij}+O\left(\alpha^{2}\right)\ , \\
\mathcal{O}&= \bar{\mathcal{O}}+\alpha\,\tilde{\mathcal{O}}+O\left(\alpha^{2}\right)\,,\\
K_{\mathcal{O}}(x,x';t) & = \bar{K}_{\mathcal{O}}(x,x';t)+\alpha~\tilde{K}_{\mathcal{O}}(x,x';t)+O\left(\alpha^{2}\right)\ , 
\end{split}
\end{equation}
where  $\bar{K}_{\mathcal{O}}$ is the  heat   kernel corresponding to $\bar{\mathcal{O}}$, i.e.
\be 
\label{orderzero}
\left(\partial_{t}+\bar{\mathcal{O}}_{x}\right)\bar{K}_{\mathcal{O}}(x,x';t) = 0\ ,   \qquad~~\bar{K}_{\mathcal{O}}(x,x';0)=\frac{1}{\sqrt{\bar{g}}}\delta^{\left(d\right)}\left(x-x'\right)\mathbb{I} \ . \ee
Then $\tilde{K}_{\mathcal{O}}$    may be found  by solving 
\be  \label{orderalpha}
\left(\partial_{t}+\bar{\mathcal{O}}_{x}\right)\tilde{K}_{\mathcal{O}}(x,x';t)+\tilde{\mathcal{O}}_{x}\bar{K}_{\mathcal{O}}(x,x';t)=0  \ , \qquad~~\tilde{K}_{\mathcal{O}}(x,x';0)=-\frac{\tilde{g}}{2\bar{g}^{3/2}}\delta^{\left(d\right)}\left(x-x'\right)\mathbb{I}\,.
\ee
 The resulting solution is (see Appendix \ref{app:heat_kernel_perturb}  for details)
\begin{eqnarray}\nonumber
\!\!\!\!
\tilde{K}_{\mathcal{O}}\left(x,x';t\right) & =&-\frac{\tilde{g}}{2\bar{g}^{3/2}}\delta^{(d)}(x-x')\mathbb{I} \no\\
& &+\int_{0}^{t}dt'\int d^d{x''}\sqrt{\bar{g}}\,\bar{K}_{\mathcal{O}}\left(x,x'';t-t'\right)\bar{\mathcal{O}}_{x''}\Big(\frac{\tilde{g}}{2\bar{g}^{3/2}}\delta^{\left(d\right)}(x''-x')\Big)\no \\\label{2.7}
 && -\int_{0}^{t}dt'\int d^d{x''}\sqrt{\bar{g}}\bar{K}_{\mathcal{O}}\left(x,x'';t-t'\right)\tilde{\mathcal{O}}_{x''}\bar{K}_{\mathcal{O}}\left(x'',x';t'\right)\,.
\end{eqnarray}
 Then the trace  $K_{\mathcal{O}}\left(t\right)$ in \rf{2.2}  takes the form 
\begin{eqnarray}\label{2.80}
&&K_{\mathcal{O}}\left(t\right) =\bar{K}_{\mathcal{O}}\left(t\right)+\alpha\,\tilde{K}_{\mathcal{O}}\left(t\right)+O\left(\alpha^{2}\right)~,\\
\label{2.8}
&&\tilde{K}_{\mathcal{O}}\left(t\right)  =-t\int d^d{x}\sqrt{\bar{g}}~~\textrm{tr}\Big[\tilde{\mathcal{O}}_{x}~\bar{K}_{\mathcal{O}}\left(x,x';t\right)\Big]_{x=x'}\,.
\end{eqnarray}
Thus the perturbative expansion of  the  determinant of $\mathcal{O}$ in \eqref{2.1}   becomes 
\begin{eqnarray}\label{det_perturb}
{ \Det_{_{\mathcal{M}}}\mathcal{O}   \over \Det_{_{\mathcal{\bar{ M}}}}\mathcal{\bar{O}}} &=& e^{ -\alpha\,\tilde\zeta_{\mathcal{O}}^{'}\left(0\right)+O(\alpha^2)} \ , 
\qquad \qquad 
\log\, \Det_{_{\mathcal{M}}}\mathcal{O}  = 
 -\bar\zeta_{\mathcal{O}}^{'}\left(0\right)-\alpha\,\tilde\zeta_{\mathcal{O}}^{'}\left(0\right)+O(\alpha^2)\,,\\\nonumber\\ \label{zeta_tilde}
\bar{\zeta}_{\mathcal{O}}\left(s\right)  &=&\frac{1}{\Gamma\left(s\right)}\int_{0}^{\infty}dt\,t^{s-1}\,{\bar K}_{\mathcal{O}}\left(t\right)\,,\qquad~~~
\tilde{\zeta}_{\mathcal{O}}\left(s\right)  =\frac{1}{\Gamma\left(s\right)}\int_{0}^{\infty}dt\,t^{s-1}\tilde{K}_{\mathcal{O}}\left(t\right)~.
\end{eqnarray}
From \rf{2.80},\rf{2.8}  one   can find similar perturbative expansions 
for the  coefficients  in the small-$t$ expansion of the heat kernel, i.e. for the Seeley coefficients  that control the UV divergent part of 
$\log\, \Det_{_{\mathcal{M}}}\mathcal{O}$  in, e.g., the proper-time  regularization (see, e.g., ~\cite{Gilkey,Fursaev}). 
As a  check of \rf{2.7}  we show in Appendix \ref{app:pert_Seeley_Laplacian} that the small-$t$ expansion of \eqref{2.8} reproduces the results of the  standard perturbation theory applied  directly to the Seeley coefficients 
of the scalar Laplace operator on a manifold with no singularities.

\iffa 
Notice that while $\Det_{_{\mathcal{\bar M}}}\mathcal{O}$ is determined by the unperturbed traced heat kernel which is by definition at coincident points,  the ratio $\log\big(\Det_{_{\mathcal{M}} \mathcal{O} /\Det_{_{\mathcal{\bar M}}}} \bar{\mathcal{O}}\big)$ is also sensible to the behaviour for $x$ close to $x'$ due to the presence of the derivatives in $\tilde{\mathcal{O}}$ in \eqref{2.8}.
\fi

 In the following section we will consider   examples where  scalar and spinor operators will be defined 
 on the two-dimensional   $\bar{\mathcal{M}}$    which will 
  be real hyperbolic space $H^2$. 
   In this case  the homogeneity of $H^2$  allows one to construct the relevant heat kernels ${\bar K}_{\mathcal{O}}$ 
   for  generic  pair of points $x,x'$
   ~\cite{Camporesi:1990wm, Camporesi:1994ga,Camporesi:1992tm,Camporesi:1995fb}  (see also Appendix \ref{app:heat_kernel_zeta_function})  and thus to compute the first corrections  ${\tilde  K}_{\mathcal{O}}$ 
   according to \rf{2.7}.

\section{Perturbative expansion of 1-loop string   correction to Wilson loop minimal surfaces }
\label{sec:applications}

Our aim will be to use the above expressions  to develop a perturbative  approach to computation of 
   $AdS_5\times S^5$ superstring   partition function $Z$ expanded near a particular minimal surface ending on the AdS boundary 
 that represents the leading  strong-coupling correction 
to  the corresponding Wilson loop in gauge theory. In general,   
\be\label{3.1}
Z= \langle W (\lambda, \alpha) \rangle \equiv e^{-\Gamma}\,,\qquad\qquad \Gamma=\sqrt{\lambda}\,
\Gamma^{(0)}(\alpha)+\Gamma^{(1)}(\alpha)+\mathcal{O}(\lambda^{-{1}/{2}}) \ .
\ee
Here $\sqrt \lambda \Gamma^{(0)}(\alpha)$ is the classical  string action ($\sqrt \lambda \over 2 \pi$ is   the string tension)
evaluated on  a minimal surface with parameter $\a$ 
and $\Gamma^{(1)}(\alpha)$ is the 1-loop    correction expressed in terms of  ratios   of determinants of 2nd order fluctuation operators 
\ci{Kallosh:1998ji,Forste:1999qn,Drukker:2000ep}. 

While computing these determinants for a generic minimal surface  is hard, expanding  in some   small parameter $\a$ 
(such that   for $\a=0$ the surface  becomes simple) 
that can be done  in perturbation theory. We shall demonstrate this   below 
 in a number of cases:

(i)   ``latitudes'' in $S^2\subset S^5$ (Section \ref{sec:latitude}); 

 (ii)   generalized  cusp   (Section \ref{sec:generalized_cusp});
 
   (iii) $k$-wound circle   (Section \ref{sec:kcircle}).  
 
 \noindent 
In these  cases the $\a=0$  limit of the  minimal surface  will 
be the Euclidean $AdS_2$  space  or $H^2$   for which  the heat kernels and determinants  
or relevant operators   are known explicitly, i.e. $\Gamma^{(1)}(0)\equiv \bar  \Gamma^{(1)}$ is known. 
Our aim will be   to  find the first correction  to $\Gamma^{(1)}(0)$:
\be \Gamma^{(1)}(\alpha) = \bar  \Gamma^{(1)} + \alpha\, \td   \Gamma^{(1)} +O\left(\alpha^{2}\right)\ . \la{3.2} 
\ee


\subsection{Latitude Wilson loop}
\label{sec:latitude}

Let us   start with a   family of 1/4-BPS  Wilson loops
 with   the minimal surface  of half-sphere topology 
  ending on a unit circle at the boundary of $AdS_5$ and   stretched also along  the  latitude  located  at the 
 polar angle $\theta_0$ in a ${{S}}^2\subset{{S}}^5$~\cite{Drukker:2005cu,Drukker:2006ga,Drukker:2007qr}. 
The minimal  surface is embedded into  a subspace $H^3\times S^2$ of $AdS_5\times S^5$ with the 
metric  \be \la{3.3} 
ds^2_{H^3\times S^2}  =  z^{-2}(dx_1^2+dx_2^2+dz^2)
 +d\theta^2+\sin^2\theta\, d\phi^2 \ee 
  as  follows 
\begin{eqnarray}\label{3.5}
\!\!\!\!\!\!\!\!\!\!\!\!
x_1&=&\frac{\cos\tau}{\cosh\sigma}\,,~~\qquad 
x_2=\frac{\sin\tau}{\cosh\sigma}\,, ~~\qquad 
z=\tanh\sigma\,,~~\\
\sin\theta&=&\frac{1}{\cosh(\sigma+\sigma_0)}\,,\qquad  ~ \cos\theta=\tanh(\sigma+\sigma_0)  \ ,\qquad 
 \phi=\tau\,, \la{3.55}
 \\\label{3.6}
& & \sigma\in[0,\infty)\,, \qquad
\tau\in [0,2\pi) \,,\qquad
\tanh\sigma_{0}\equiv\cos\theta_{0}\,. 
\end{eqnarray}
The world-sheet boundary at
$\sigma=0$ is located at the boundary of $AdS_5$, and $\sigma_0\in[0,\infty)$ related to $\theta_0 \in[0,\frac{\pi}{2}]$ 
describes a    one-parameter family of latitudes on $S^5$. 
The  maximally supersymmetric (1/2-BPS) case corresponds to  $\theta_0=0$ or $\sigma_0 =\infty$ 
when  the latitude in $S^2$   shrinks to a point  ($\theta=\theta_0=0$) 
 and thus  the minimal  surface  becomes  the same as of the circular Wilson loop. 
 In what   follows $\theta_0$ will thus play the role of the small expansion parameter $\a$. 

The   induced world-sheet  geometry     is  that of   the  2d   Euclidean manifold $\mathcal{M}$ with  the metric
\be\label{la_metric}
\begin{split}
ds^{2}_{\mathcal{M}}&
=\Omega^{2}\left(\sigma\right)\left(d\tau^{2} + d\sigma^{2}\right)\ , 
\\
\Omega^{2}\left(\sigma\right)&\equiv\frac{1}{\sinh^{2}\sigma}+\frac{1}{\cosh^{2}\left(\sigma+\sigma_{0}\right)}=\frac{1}{\sinh^{2}\sigma}+O\left(\theta_{0}^{2}\right)\ ,
\end{split}
\ee
which for $\s_0 =\infty$, i.e. $\theta_{0}=0$,  becomes  the  hyperbolic plane $H^{2}$.  
The  leading term in \rf{3.1}, i.e. the 
 area of this minimal  surface,  regularized in a standard way  by 
 introducing a small cutoff near the boundary of $AdS_5$, at $z=\epsilon\to 0$,  or, 
  equivalently,  at  $\sigma=\arctanh\, \epsilon\to \infty $ is then
\be\label{3.8}
\Gamma^{(0)}(\theta_{0})
= \frac{1}{2\pi} \int_{0}^{2\pi} d\tau\int_{\arctanh\, \epsilon}^{\infty}d\sigma \,\, \Omega^2(\sigma) 
=\frac{1}{\epsilon}-\cos\theta_{0}
\to- \cos\theta_{0}\,.
\ee
The  singular  term  here is $\theta_0$-independent   and thus is  the same as in the singular part of 
 the volume of  Euclidean $AdS_2$ space.\foot{The
   linearly divergent part $1\ov \epsilon$, proportional to the length of the  boundary at $z=\epsilon$,   may be 
    subtracted by a Legendre transform of the  Wilson loop  as in ~\cite{Drukker:1999zq,Drukker:2000ep,Drukker:2005kx}.}

Expanding the   $AdS_5 \times S^5$ 
 superstring action  to second order in the fluctuation fields 
leads to  the following one-loop contribution to \rf{3.1} ~\cite{Forini:2015mca,Forini:2015bgo,Faraggi:2016ekd}~\footnote{As in  earlier discussions \ci{Drukker:2000ep,Forini:2015mca}  it is   assumed   here 
 that the same boundary conditions are imposed on the  operator 
 of the longitudinal bosonic modes and the one of the ghosts associated with  the diffeomorphisms gauge-fixing, so that their 
net contribution to the ratio \eqref{3.9}  equals to one.}
\begin{gather}\label{3.9}
\Gamma^{(1)}\left(\theta_{0}\right)=-\log\frac{\prod_{p_{12},p_{56}=\pm1}\textrm{Det}^{2/4}\Big[\mathcal{O}_{p_{12},p_{56}}^{2}(\theta_{0})\Big]}{\textrm{Det}^{3/2}\Big[\mathcal{O}_{1}(\theta_0)\Big]\,\textrm{Det}^{3/2}\Big[\mathcal{O}_{2}(\theta_{0})\Big]\,\textrm{Det}^{1/2}\Big[\mathcal{O}_{3+}(\theta_{0})\Big]\,\textrm{Det}^{1/2}\Big[\mathcal{O}_{3-}(\theta_{0})\Big]}\,.
\end{gather}
Here the bosonic second-order operators~\footnote{The operators $\mathcal{O}_{3\pm}(\theta_0)$ in (3.10) of~\cite{Forini:2015bgo} coincide with the ones in \eqref{lbo_O3pm}  upon the replacement $-i \partial_\tau \to -i \partial_\tau \pm 1$, which implements the shift explained in Section 4 of~\cite{Forini:2015bgo}. This is equivalent to a choice of the normal bundle gauge connection~\cite{Forini:2015mca} that is regular everywhere on the world-sheet (see discussion below (4.20) of~\cite{Faraggi:2016ekd}).}
\begin{flalign}\label{lbo_O1}
\mathcal{O}_{1}(\theta_0) & \equiv\frac{1}{\Omega^{2}(\sigma)}\Big(-\partial_{\tau}^{2}-\partial_{\sigma}^{2}+\frac{2}{\sinh^{2}\sigma}\Big)\ , \quad
\mathcal{O}_{2}\left(\theta_{0}\right)  \equiv\frac{1}{\Omega^{2}(\sigma)}\Big(-\partial_{\tau}^{2}-\partial_{\sigma}^{2}-\frac{2}{\cosh^{2}\left(\sigma+\sigma_{0}\right)}\Big)\ , \\
\no 
\mathcal{O}_{3\pm}\left(\theta_{0}\right) & \equiv\frac{1}{\Omega^{2}(\sigma)}\Big[-\partial_{\tau}^{2}-\partial_{\sigma}^{2}\pm2i\left(\tanh\left(2\sigma+\sigma_{0}\right)-1\right)\partial_{\tau} \\
& \qquad \qquad ~~  -1-2\tanh\left(2\sigma+\sigma_{0}\right)+3\tanh^{2}\left(2\sigma+\sigma_{0}\right)\Big]\label{lbo_O3pm}
\end{flalign}
act on the world-sheet scalars,  and the fermionic  first-order operators  
\begin{flalign}\label{lbf}
\mathcal{O}_{p_{12},p_{56}}\left(\theta_{0}\right) & \equiv
\frac{i}{\Omega(\sigma)}\Big(\partial_{\sigma}+\frac{\Omega{'}(\sigma)}{2\,\Omega(\sigma)}\Big)\sigma_{1}
+\frac{1}{\Omega(\sigma)}\left(-i\partial_{\tau}+\frac{p_{56}}{2}\big[1-\tanh\left(2\sigma+\sigma_{0}\right)\big] \right)\sigma_{2}\no\\
&\quad  +\frac{p_{12}}{\Omega^{2}(\sigma)\sinh^{2}\sigma}\sigma_{3} -\frac{p_{12}\,p_{56}}{\Omega^{2}(\sigma)\cosh^{2}\left(\sigma+\sigma_{0}\right)}\mathbb{I}_{2}
\end{flalign}
act on two-dimensional spinors and are labeled by $p_{12},p_{56}=\pm 1$ ($\sigma_i$ are Pauli matrices).\footnote{Compared to the notation used in (3.26) of~\cite{Forini:2015bgo}, the fermionic determinants are raised  in \eqref{3.9} 
to an additional power of two because the irrelevant label $p_{89}$ is suppressed. 
We also made the replacement $-i\partial_\tau\to -i\partial_\tau +\frac{p_{56}}{2}$ to arrive at \eqref{lbf}, as motivated in Section 4 of~\cite{Forini:2015bgo}. }
The determinants of these operators have been evaluated exactly (for any $\theta_0$)  in~\cite{Forini:2015bgo,Faraggi:2016ekd}.

To   apply  the perturbative approach developed in Section \ref{sec:perturbative_hk}, we choose 
\be
\label{3.14} 
\alpha_{_{\rm latitude}}\equiv \theta^2_0\,,
\ee
so that the  reference manifold $\bar{\mathcal{M}}$  for $\a=0$ is 
 $H^2$  corresponding  to the circular Wilson loop ($\theta_0=0$, or $\sigma_0=\infty)$, i.e. 
\begin{gather}\label{3.15}
ds^{2}_{\bar{\mathcal{M}}}=\frac{d\tau^{2}+d\sigma^{2}}{\sinh^{2}\sigma}=d\rho^{2}+\sinh^{2}\rho\, d\tau^{2}\,,
\qquad\qquad
\sinh\rho\equiv\frac{1}{\sinh\sigma}\,,
\end{gather}
with  the $S^1$ boundary at  
\be \la{3.155}
 \rho=\ve \to \infty \ , \ \ \ \ \ \ \ \       \ve \equiv \textrm{arccosh}(\epsilon^{-1}) \ . \ee
The   string action  proportional to the  (renormalized) volume  of this space is  
\begin{gather}\label{3.17}
\Gamma^{(0)}({0}) = { 1 \over 2 \pi}  V_{H^{2}}=
  { 1 \over 2 \pi} \int_{0}^{2\pi}d\tau\int_{0}^\ve d\rho\:\sinh\rho
=\frac{1}{\epsilon}-1  \to- 1 \,, 
\end{gather}
which is the $\theta_0=0$ term in \rf{3.8}.
In the limit $\theta_0=0$ the operators \eqref{lbo_O1}-\eqref{lbf} take the form of the Laplacian \eqref{polar_Laplace_operator} and the Dirac operator \eqref{polar_Dirac_operator}
\begin{gather}\label{la_operator_order_0}
\bar{\mathcal{O}}_{1}  =-\Delta_{\rho,\tau}+2\,,\qquad
\bar{\mathcal{O}}_{2}=\bar{\mathcal{O}}_{3\pm}  =-\Delta_{\rho,\tau}\,,\qquad
\bar{\mathcal{O}}_{p_{12},p_{56}} =-i\slashed{\nabla}_{\rho,\tau}+p_{12}\,\sigma_{3}\,.
\end{gather}
The spectrum of physical excitations  which contribute to $\Gamma^{(1)}\left(\theta_{0}=0\right)$ in \eqref{3.9},
 is composed of 3 massive scalars
$\left(m^{2}=2\right)$, 5 massless scalars and 8 massive 2d Majorana
spinors ($m^{2}=1$) propagating in $H^2$~\cite{Drukker:2000ep,Kruczenski:2008zk}. The regularized determinants were computed in~\cite{Buchbinder:2014nia} with the heat kernel method using  \eqref{ze_scalar} and \eqref{ze_spinor}
\begin{flalign}
\bar{\zeta}'_{\mathcal{O}_{1}}(0)&=-\frac{25}{12}+\frac{3}{2}\log2\pi-2\log A\,,
\\
\bar{\zeta}'_{\mathcal{O}_{2}}(0)=\bar{\zeta}'_{\mathcal{O}_{3\pm}}(0)&=-\frac{1}{12}+\frac{1}{2}\log2\pi-2\log A\,,
\\
\bar{\zeta}'_{\mathcal{O}_{p_{12},p_{56}}^{2}}(0)&=-\frac{5}{3}+2\log2\pi-4\log A\,,
\end{flalign}
where $A$ is the Glaisher constant  (see \eqref{formula3}, \eqref{formula4} and \eqref{formula5}). 
As  a  result,  
 the one-loop correction \eqref{3.9}  in the  circular   Wilson loop  case is 
\be\label{3.9_order_0}
\Gamma^{(1)}(0)  = -\frac{3}{2}\bar{\zeta}^{'}_{\mathcal{O}_{1}}(0)-\frac{3}{2}\bar{\zeta}^{'}_{\mathcal{O}_{2}}(0)-\frac{1}{2}\bar{\zeta}^{'}_{\mathcal{O}_{3+}}(0)-\frac{1}{2}\bar{\zeta}_{\mathcal{O}_{3-}}^{'}(0)+\frac{1}{2}\!\!\!\!\!\!
\sum_{p_{12},p_{56}=\pm1}\bar{\zeta}^{'}_{\mathcal{O}_{p_{12},p_{56}}^{2}}(0)= \frac{1}{2}\log2\pi.
\ee
Expanding  \rf{la_metric}  in small $\a= \theta_0^2$   we find   that  the  leading  correction to the metric   \rf{3.15} 
 in \rf{expansion}  is given by  
 \begin{flalign}
\bar{g}_{ij}(\rho,\tau)&=
\left(\begin{array}{cc}
1 & 0\\
0 & \sinh^{2}\rho
\end{array}\right)\,,
\qquad
\tilde{g}_{ij}(\rho,\tau)=
\left(\begin{array}{cc}
\frac{1}{(1+\cosh\rho)^{2}} & 0\\
0 & \frac{\cosh\rho-1}{\cosh\rho+1}
\end{array}\right)\,.
\end{flalign}
From \eqref{lbo_O1}-\eqref{lbf} we  find 
that the  expansion of   the relevant differential operators~\footnote{Here by  $\left\{,\right\}$ we indicate  the anticommutator of two (matrix-valued) differential operators.} 
\begin{flalign}\label{lbo_O1_expansion}
\mathcal{O}_{i}(\theta_0) & =\bar{\mathcal{O}}_{i}+\theta_{0}^{2}\,\tilde{\mathcal{O}}_{i}+O(\theta_0^4)\,,
\qquad\qquad i=1,\,2,\,3+,\,3-\,,\\
\label{lbf_expansion}
\mathcal{O}_{p_{12},p_{56}}(\theta_0) & =\bar{\mathcal{O}}_{p_{12},p_{56}}+\theta_{0}^{2}\,\tilde{\mathcal{O}}_{p_{12},p_{56}}+O(\theta_0^4)\,,\\
\label{lbf_squared_expansion}
\mathcal{O}_{p_{12},p_{56}}^{2}(\theta_0) &  =\bar{\mathcal{O}}_{p_{12},p_{56}}^{2}+\theta_{0}^{2}\,\left\{ \bar{\mathcal{O}}_{p_{12},p_{56}},\tilde{\mathcal{O}}_{p_{12},p_{56}}\right\} +O(\theta_0^4)\ , 
\end{flalign}
contains 
\begin{flalign}\label{lbo_O1_order_1}
\tilde{\mathcal{O}}_{1}=\tilde{\mathcal{O}}_{2} & =\frac{1}{\left(1+\cosh\rho\right)^{2}}\left(\Delta_{\rho,\tau}-2\right)\,,\\
\label{lbo_O3pm_order_1}
\tilde{\mathcal{O}}_{3\pm} & =\frac{1}{\left(1+\cosh\rho\right)^{2}}\Big[\Delta_{\rho,\tau}-\frac{\sinh^2\rho}{(1+\cosh\rho)^2}\left(2\pm i\partial_{\tau}\right)\Big]\,,\\
\label{lbf_order_1}
\tilde{\mathcal{O}}_{p_{12},p_{56}} & =\frac{i}{2\left(1+\cosh\rho\right)^{2}}\slashed{\nabla}_{\rho,\tau}-\frac{i\left(1-\cosh\rho\right)}{2\sinh\rho\left(1+\cosh\rho\right)^{2}}\sigma_{1}\,\no  \\
& +\frac{p_{56}\sinh^{3}\rho}{4\left(1+\cosh\rho\right)^{4}}\sigma_{2}-\frac{p_{12}}{\left(1+\cosh\rho\right)^{2}} \left(\sigma_{3}+p_{56}\,\mathbb{I}_{2}\right)\,.
\end{flalign}
For the bosonic operator $\mathcal{O}_1(\theta_0)$ in \eqref{lbo_O1_expansion},  
 substituting  \eqref{lbo_O1_order_1} into \eqref{2.8}, we obtain
\begin{gather}\label{fe_1}
\!\!
\tilde{K}_{\mathcal{O}_{1}}\left(t\right) =-t \int_{0}^{2\pi} 
 d\tau \int_{0}^\ve
 d\rho \frac{\sinh\rho}{\left(1+\cosh\rho\right)^{2}} \Big[\left(\Delta_{\rho,\tau}-2\right)\bar{K}_{-\Delta+2}(\rho,\tau,\rho',\tau';t)\Big]_{\rho=\rho',\tau=\tau'}\,,
\end{gather}
where $\ve$ was defined in \rf{3.155}. 
 As $\bar{\mathcal{O}}_1$ in \eqref{la_operator_order_0} is the Laplacian for a scalar field of mass $m^2=2$,  its heat 
 kernel satisfies    
\be\label{fe_2}
\left(\partial_t-\Delta_{\rho,\tau}+2\right)\bar{K}_{\mathcal{O}_1}(\rho,\tau,\rho',\tau';t)=0\,
\ee
so 
that we can trade the Laplacian in \eqref{fe_1}  for  the derivative $\partial_t$, and then take the coincident-point limit, getting 
\begin{gather}\label{fe_3}
\tilde{K}_{\mathcal{O}_{1}}\left(t\right) =-t \int_{0}^{2\pi} d\tau \int_{0}^\ve d\rho \frac{\sinh\rho}{\left(1+\cosh\rho\right)^{2}}\,\partial_t \bar{K}_{\mathcal{O}_1}\left(\rho,\tau,\rho,\tau;t\right)\,.
\end{gather}
Here   we can send  the upper limit to infinity ($\ve \to \infty$  corresponds to $\epsilon\to 0$  in \rf{3.155}) 
and then use the integral representation of the traced heat kernel \eqref{traced_hk_Laplace} for mass $m^2=2$
\begin{gather}\label{fe_4}
\tilde{K}_{\mathcal{O}_{1}}\left(t\right) =\frac{t}{2}\int_{0}^{\infty}dv\,v\tanh\left(\pi v\right)
\te  \big(v^{2}+\frac{9}{4}\big)\ e^{-t\left(v^{2}+\frac{9}{4}\right)}\,.
\end{gather}
To evaluate $\tilde{\zeta}_{\mathcal{O}_{1}}(s)$ one proceeds as in  Appendix \ref{app:zeta_functions}, 
interchanging the integration over the spectral parameter $v$ and the proper time $t$ in the definition \eqref{zeta_tilde} of the zeta-function, and writing $\tanh(\pi v) =1-{2}/({e^{2\pi v}+1})$ to get
\be\label{fe_5}
\tilde{\zeta}_{\mathcal{O}_{1}}(s) 
= \int_{0}^{\infty}dv\frac{sv}{2\left(v^{2}+\frac{9}{4}\right)^{s}}- \int_{0}^{\infty}dv\frac{sv}{(e^{2\pi v}+1)\left(v^{2}+\frac{9}{4}\right)^{s}}\\
\ . \ee
%
As the first integral above converges only for $\textrm{Re}\,s>1$, one  can  first integrate over 
 $v$  assuming this is true  and then  analytically continue to all  values of $s$
\begin{gather}\label{fe_6}
\tilde{\zeta}_{\mathcal{O}_{1}}\left(s\right) ={\te 
\frac{s}{4\left(s-1\right)}{ \left(\frac{9}{4}\right)^{1-s}} } -s\int_{0}^{\infty}dv\frac{v}{\left(e^{2\pi v}+1\right)\left(v^{2}+\frac{9}{4}\right)^{s}}\,, 
\end{gather}
and one obtains
\begin{flalign}\label{ze_O1_order_1}
\tilde{\zeta}_{\mathcal{O}_{1}}^{'}\left(0\right) =\te -\frac{7}{12}\,.
\end{flalign}
The same steps may be followed for 
$\mathcal{O}_2(\theta_0)$, for which one gets 
\begin{flalign}
\label{fe_18}
\tilde{K}_{\mathcal{O}_{2}}\left(t\right) 
 & =\frac{t}{2}\int_{0}^{\infty}dv\,v\tanh\left(\pi v\right){{\te \left(v^{2}+\frac{9}{4}\right)}} \ e^{-t\left(v^{2}+\frac{1}{4}\right)} \ , 
\\
\no 
%
%
\tilde{\zeta}_{\mathcal{O}_{2}}(s) 
&=\te  \int_{0}^{\infty}dv {\te 
\frac{sv}{\left(v^{2}+\frac{1}{4}\right)^{s}}\Big(\frac{1}{2}+\frac{1}{v^{2}+\frac{1}{4}}\Big)}
- \int_{0}^{\infty}dv{\te  \frac{sv}{(e^{2\pi v}+1)\left(v^{2}+\frac{1}{4}\right)^{s}}\Big(1+\frac{2}{v^{2}+\frac{1}{4}}\Big)}
 \no
  \\
 & ={\te {\frac{s}{4\left(s-1\right)}\left(\frac{1}{4}\right)^{1-s}+\frac{1}{2}\left(\frac{1}{4}\right)^{-s}  }
 -{s}\int_{0}^{\infty}dv\frac{v}{\left(e^{2\pi v}+1\right)\left(v^{2}+\frac{1}{4}\right)^{s}}
-2s\int_{0}^{\infty}dv\frac{v}{\left(e^{2\pi v}+1\right)\left(v^{2}+\frac{1}{4}\right)^{s+1}}}\ ,\no\\
\label{ze_O2_order_1}
\tilde{\zeta}_{\mathcal{O}_{2}}^{'}\left(0\right) &=\te  -\frac{1}{12}+{\gamma}\,.
\end{flalign}
Here  we used   \eqref{formula7}  and  $\gamma$ is the Euler-Mascheroni constant.

The operators  $\mathcal{O}_{3+}(\theta_0)$ and $\mathcal{O}_{3-}(\theta_0)$  coincide for $\theta_0=0$ in \eqref{la_operator_order_0} and therefore 
 the derivatives $\partial_\tau$ in \eqref{lbo_O3pm_order_1} cancel each other in the sum~\footnote{The derivatives come with opposite signs in~\eqref{lbo_O3pm_order_1}  as the fields acted upon by \eqref{lbo_O3pm} in the  fluctuation Lagrangian~\cite{Forini:2015mca,Forini:2015bgo,Faraggi:2016ekd}  are a complex scalar and its complex conjugate, coupled to  a $U(1)$   connection   with opposite charges~\cite{Faraggi:2016ekd}.  
}
\begin{flalign}
 &\tilde{K}_{\mathcal{O}_{3+}}\left(t\right)+\tilde{K}_{\mathcal{O}_{3-}}\left(t\right)   \no \\ 
& ={-2t \int_{0}^{2\pi} d\tau \int_{0}^\infty d\rho \frac{\sinh\rho}{(1+\cosh\rho)^{2}}
\Big[\Big(\Delta_{\rho,\tau}-2\frac{\sinh^2\rho}{(1+\cosh\rho)^2}\Big)\bar{K}_{-\Delta}(\rho,\tau,\rho',\tau';t)\Big]_{\rho=\rho',\tau=\tau'}}\no\\
\label{fe_19}
 & ={t}\int_{0}^{\infty}dv\,v\tanh\left(\pi v\right){\te \left(v^{2}+\frac{5}{4}\right)} \ e^{-t\big(v^{2}+\frac{1}{4}\big)}\,.
\end{flalign}
Then for the combined zeta-functions one obtains
\begin{flalign}\label{fe_10}
%
%
& \tilde{\zeta}_{\mathcal{O}_{3+}}(s)+\tilde{\zeta}_{\mathcal{O}_{3-}}(s) 
 =\textstyle \int_{0}^{\infty}dv\frac{sv}{(v^{2}+\frac{1}{4})^{s}}(1+\frac{1}{v^{2}+\frac{1}{4}})+\int_{0}^{\infty}dv\frac{-2sv}{(e^{2\pi v}+1)(v^{2}+\frac{1}{4})^{s}}(1+\frac{1}{v^{2}+\frac{1}{4}}) \\
 &~~~~~~ =\textstyle{\frac{s}{2\left(s-1\right)}\left(\frac{1}{4}\right)^{1-s}+\frac{1}{2}\left(\frac{1}{4}\right)^{-s}
 -2s\int_{0}^{\infty}dv\frac{v}{\left(e^{2\pi v}+1\right)\left(v^{2}+\frac{1}{4}\right)^{s}}-2s\int_{0}^{\infty}dv\frac{v}{\left(e^{2\pi v}+1\right)\left(v^{2}+\frac{1}{4}\right)^{s+1}}}\no\,,\\
\label{ze_O3pm_order_1}
& \tilde{\zeta}_{\mathcal{O}_{3+}}^{'}\left(0\right)+\tilde{\zeta}_{\mathcal{O}_{3-}}^{'}\left(0\right) =\te  -\frac{1}{6}+{\gamma}\,,
\end{flalign}
where we used \eqref{formula7}. 
In the fermionic case the 
 relevant operator is the \emph{square} of $\mathcal{O}_{p_{12},p_{56}}(\theta_0)$,  a positive-definite operator with a well-defined $\theta_0$-expansion of its heat kernel defined in 
\eqref{lbf_expansion} 
\begin{flalign}
\tilde{K}_{\mathcal{O}_{p_{12},p_{56}}^{2}}(t) & =-t 
\int_{0}^{2\pi} d\tau \int_{0}^\infty   d\rho\, \sinh\rho\, 
\textrm{tr} \Big[\Big\{ \bar{\mathcal{O}}_{p_{12},p_{56}}^{\rho,\tau},\tilde{\mathcal{O}}_{p_{12},p_{56}}^{\rho,\tau}\Big\} \bar{K}_{-\slashed{\nabla}^{2}+1}(\rho,\tau,\rho{'},\tau{'};t)\Big]_{\rho=\rho{'},\tau=\tau{'}}\no\\
\label{fe_7}
 & =t\int_{0}^{\infty}dv\,v\coth\left(\pi v\right)\left(v^{2}+2\right)e^{-t\left(v^{2}+1\right)}\,.
\end{flalign}
Here one has to work with the full  heat kernel  \eqref{b13} for $m^2=1$  and the  rest of the computation is essentially unchanged, giving 
\begin{flalign} \no   
\!\!\!\!\!\!\!\!
\tilde{\zeta}_{\mathcal{O}_{p_{12},p_{56}}^{2}}\left(s\right) 
 &=\textstyle \int_{0}^{\infty}dv\frac{sv}{\left(v^{2}+1\right)^{s}}(1+\frac{1}{v^{2}+1})+\int_{0}^{\infty}dv\frac{2sv}{\left(e^{2\pi v}-1)(v^{2}+1\right)^{s}}(1+\frac{1}{v^{2}+1})  \\
\!\!\!\!\!\!\!\!
 &=\textstyle{\frac{s}{2(s-1)}+\frac{1}{2}+2s\int_{0}^{\infty}dv\frac{v}{\left(e^{2\pi v}-1\right)\left(v^{2}+1\right)^{s}}
+2s\int_{0}^{\infty}dv\frac{v}{\left(e^{2\pi v}-1\right)\left(v^{2}+1\right)^{s+1}}}\la{fe_11}  \ , \\
\label{ze_OF_order_1}
\!\!\!\!\!\!\!\!
\tilde{\zeta}_{\mathcal{O}_{p_{12},p_{56}}^{2}}^{'}\left(0\right) & =\te -\frac{11}{12}+\gamma \,.
\end{flalign}
where we split $\coth(\pi v) =1+{2}/({e^{2\pi v}-1})$ and the last relation  follows from \eqref{formula8}.

We can now sum over the bosonic and fermionic contributions  to get 
\begin{flalign}\la{3.45}
& \Gamma^{(1)}\left(\theta_{0}\right)- \Gamma^{(1)}(0) =  \theta_0^2 \, \td \Gamma^{(1)}  +O(\theta_0^4) \ ,  \\
\qquad   \td \Gamma^{(1)} &=\te -\frac{3}{2}\tilde{\zeta}^{'}_{\mathcal{O}_{1}}(0)-\frac{3}{2}\tilde{\zeta}^{'}_{\mathcal{O}_{2}}(0)-\frac{1}{2}\tilde{\zeta}^{'}_{\mathcal{O}_{3+}}(0)-\frac{1}{2}\tilde{\zeta}_{\mathcal{O}_{3-}}^{'}(0)
+\frac{1}{2}\sum_{p_{12},p_{56}=\pm1}\tilde{\zeta}^{'}_{\mathcal{O}_{p_{12},p_{56}}^{2}}(0) \no 
\\ \qquad  &  = \te -\frac{3}{4}\ .
\label{3.46}
\end{flalign}
Remarkably, we thus    find  the 
 agreement 
 with the   strong-coupling expansion of  the  exact gauge-theory  result \eqref{1.1}, expanded  also in small 
  $\theta_0$. 

Let  us note 
 that to the same result \eqref{3.46}   can be found   by  reversing  the order of taking 
 the derivative in the zeta-function variable $s$ and   summing  over the scalar and spinor fields.
  The expressions  for zeta-functions in  \eqref{fe_5}--\eqref{fe_11} above 
are  written as
 $\tilde{\zeta}_{\mathcal{O}}(s)\equiv\tilde{\zeta}_{\mathcal{O}}^{(\textrm{power})}(s)+\tilde{\zeta}_{\mathcal{O}}^{(\textrm{exp})}(s)$, 
where $\tilde{\zeta}_{\mathcal{O}}^{(\textrm{power})}(s)$  includes the $1$ from the expansion of the hyperbolic functions and is defined for $\textrm{Re}\,s>1$, and $\tilde{\zeta}_{\mathcal{O}}^{(\textrm{exp})}(s)$ is well-defined for~$s$ close to~$0$.  The  analytic continuation of each $\tilde{\zeta}_{\mathcal{O}}^{(\textrm{power})}(s)$ is not necessary if one considers, before taking the derivative, the sum of all (perturbed) zeta-functions. It can be easily checked that the sum of ``power" contributions
\begin{flalign}\label{3.466}
 &\te  \frac{3}{2}\tilde{\zeta}_{\mathcal{O}_{1}}^{(\textrm{power})}(s)+\frac{3}{2}\tilde{\zeta}_{\mathcal{O}_{2}}^{(\textrm{power})}(s)+\frac{1}{2}\tilde{\zeta}_{\mathcal{O}_{3}}^{(\textrm{power})}(s)-\frac{1}{2}\sum_{p_{12},p_{56}=\pm1}\tilde{\zeta}_{\mathcal{O}_{p_{12},p_{56}}^{2}}^{(\textrm{power})}(s)\\
 &=\textstyle \int_{0}^{\infty}dv\big [\frac{3sv}{4\left(v^{2}+\frac{9}{4}\right)^{s}}+\frac{3sv}{2\left(v^{2}+\frac{1}{4}\right)^{s}}(\frac{1}{2}+\frac{1}{v^{2}+\frac{1}{4}}) 
 +\frac{sv}{2\left(v^{2}+\frac{1}{4}\right)^{s}}(1+\frac{1}{v^{2}+\frac{1}{4}})-
\frac{2sv}{\left(v^{2}+1\right)^{s}}(1+\frac{1}{v^{2}+1}) \big]\,  \no
\end{flalign}
is well defined for $\textrm{Re}\,s>s_0$ for a certain negative $s_0$. One may then  first take $s$-derivative of  the \emph{integrands} in 
\be\label{zeta_tot_latitude}
\tilde\zeta_{\rm tot}(s)=\te  \frac{3}{2}\tilde{\zeta}_{\mathcal{O}_{1}}(s)+\frac{3}{2}\tilde{\zeta}_{\mathcal{O}_{2}}(s)+\frac{1}{2}\tilde{\zeta}_{\mathcal{O}_{3+}}(s)+\frac{1}{2}\tilde{\zeta}_{\mathcal{O}_{3-}}(s)\,
-\frac{1}{2}\sum_{p_{12},p_{56}=\pm1}\tilde{\zeta}_{\mathcal{O}_{p_{12},p_{56}}^{2}}(s)\ ,
\ee
set $s=0$ and then integrate over $v$. 
It is easy to check that this leads  again to  \eqref{3.46}. 

One may track down the origin of such regular behavior for the full sum \eqref{zeta_tot_latitude}  by studying   
   the small-$t$ expansion of the   leading correction terms in  heat kernels in  \eqref{fe_4}, \eqref{fe_18}, \eqref{fe_19}, \eqref{fe_7}.  For  that one
   may   isolate  the exponentials of  $t$  and integrate the rest \footnote{Equivalently, as explained in Appendix \ref{app:pert_Seeley_Laplacian}, one could use \eqref{a_tilde}.}, 
\ba\no 
\tilde{K}_{\mathcal{O}_{1}}(t) &=&\te \frac{t}{2}\int_{0}^{\infty}dv\ (v-\frac{2v}{e^{2\pi v}+1})(v^{2}+\frac{9}{4})e^{-t(v^{2}+\frac{9}{4})} \\
 &=&\te \frac{4+9t}{16t}e^{-9t/4}-{t}\int_{0}^{\infty}dv\frac{v}{e^{2\pi v}+1} (v^{2}+\frac{9}{4})e^{-t(v^{2}+\frac{9}{4})}= \frac{1}{4t}+O(t)\,, 
 \no 
\\  \no 
\tilde{K}_{\mathcal{O}_{2}}(t) &=&\te \frac{1}{4t}+\frac{1}{2}+O(t)\,,
\qquad
\tilde{K}_{\mathcal{O}_{3+}}(t)+\tilde{K}_{\mathcal{O}_{3-}}(t) \te = \frac{1}{2t}+\frac{1}{2}+O(t)\,,\\
\label{fe_23}
\tilde{K}_{\mathcal{O}_{p_{12},p_{56}}^{2}}(t)\te & =&\te \frac{1}{2\,t}+\frac{1}{2}+O(t) \,.
\ea
Then considering the  zeta-function 
\be
\tilde{\zeta}_{\mathcal{O}}(s)=
\frac{1}{\Gamma(s)}\int_{0}^{1}dt\,t^{s-1}\tilde{K}_{\mathcal{O}}(t)
+
\frac{1}{\Gamma(s)}\int_{1}^{\infty}dt\,t^{s-1}\tilde{K}_{\mathcal{O}}(t)\,
\ee
one  finds  that the second integral  here  is finite for $s= 0$
 while  the first one  is singular due to the asymptotics in \eqref{fe_23} \footnote{More generally, since the operators \eqref{lbo_O1}--\eqref{lbo_O3pm} and the square of \eqref{lbf} have positive eigenvalues, the Mellin transform of their heat kernel traces \eqref{zeta_tilde} is convergent at the upper limit of the integral and singularities  originate only from $t=0$  (cf. \cite{Vassilevich:2003xt,Fursaev}).}.
 This  explains the need to  analytically extend   zeta-functions  to $s=0$ before computing their derivatives. 

The $t\to 0$ singularities    cancel in the sum of heat traces, due to the  special 
 spectrum  of scalar and spinor fields  and the values of their masses  
\be\label{fe_322}
\!\!\! {\te 
\frac{3}{2}\tilde{K}_{\mathcal{O}_{1}}(t)+\frac{3}{2}\tilde{K}_{\mathcal{O}_{2}}(t)+\frac{1}{2}\tilde{K}_{\mathcal{O}_{3+}}(t)+\frac{1}{2}\tilde{K}_{\mathcal{O}_{3-}}(t)-\frac{1}{2}}\sum_{p_{12},p_{56}=\pm1}\tilde{K}_{\mathcal{O}_{p_{12},p_{56}}^{2}}(t)=\frac{0}{t}+0+O(t)\,.
\ee
Thus,    in the $\theta^2_0$ term in  the  total  zeta-function \eqref{zeta_tot_latitude} 
no analytic continuation to $s=0$ is necessary. 
This  regularity  of the    leading   correction \rf{fe_322}  to the sum of traces of heat kernels  
or, equivalently, the UV   finiteness of the $\theta^2_0$  term (and, in fact, higher   terms)    in  the expansion  of the logarithm of the string 
1-loop partition   function has a simple explanation.
 The  logarithmic UV  divergences  (determined   by  the  Seeley coefficient $a_2$ of the $t^0$ part 
 in the small-$t$ expansion of heat kernel)
  in 2d are proportional,  for smooth manifolds,  to the Euler number 
which is the same  for both  the minimal surface \rf{la_metric}  and its $\theta_0 =0$ limit \rf{3.15},  
both  having the  same topology  (see also~\cite{Forini:2015bgo}).\foot{\label{footnote_divergences} The 
part of the 1-loop superstring partition function on the disc    given by the ratio of determinants as in \rf{3.9} 
is known to contain a universal    logarithmic UV divergence  which is   cancelled in the total partition function  against   the 
cutoff dependent factors in the conformal Killing
vector measure  included ~\cite{Drukker:2000ep}.}
These divergences thus cancel in the  ratio of the partition functions  of the latitude   and the  circle  minimal surfaces, i.e. in 
$\Gamma (\theta_0) - \Gamma(0)$.


\subsection{
Cusped Wilson loop      }
\label{sec:generalized_cusp}

 Next, let us   consider the string world-sheet ending on a pair of  oppositely   oriented (``antiparallel") 
  lines in  $\mathbb{R}\times S^3 \subset AdS_5$, 
 separated by a geometric angle $\pi-\phi$ along a great circle of $S^3$  (that can be   mapped   to a cusp on the plane) 
  and with  an  internal (R-symmetry) angle~$\theta$. 
 The classical solution was written in~\cite{Drukker:2007qr} 
  in terms of Jacobi elliptic functions~\footnote{We adhere to the notation in Appendix F of~\cite{Drukker:2011za}: $\textrm{sn}, \textrm{cn}, \textrm{dn}$ are the three basic Jacobi elliptic functions, $\mathbb{K}$ is the complete elliptic integral of the first kind and $\Pi$ is the complete elliptic integral of the third kind.}. Here we  will consider   only 
   the case of vanishing $\theta$~\cite{Drukker:1999zq, Forini:2010ek}. Then 
    the angular opening $\phi$ and the parameters $b,p,q$ of the classical solution in Appendix B of~\cite{Drukker:2011za}
    can   be  expressed in terms of just one  independent parameter  $k\in[0\,,\frac{1}{\sqrt{2}})$
\begin{gather}\label{lines_phi}
\!\!\!\! \te 
b=\frac{\sqrt{1-2k^{2}}}{k}\,,
~~~
 p^2=\frac{b^{4}}{1+b^{2}}\,,
~~~
q=0\,,~~~\phi =\pi-\frac{2p^{2}}{b\sqrt{b^{4}+p^{2}}}\Big[\Pi\big(\textstyle{\frac{b^{4}}{b^{4}+p^{2}}|k^{2}}\big)-\mathbb{K}\big(k^{2}\big)\Big]
\end{gather}
 and the  classical surface $\mathcal{M}$ lies entirely inside an $AdS_3$ subspace of $AdS_5$ with the metric $ds^2_{AdS_3}  =  -\cosh^2\rho\, dt^2+d\rho^2 +\sinh^2\rho\, d\varphi^2 $. After  $t\to i t $, the induced world-sheet metric is Euclidean
\begin{gather}\label{lines_metric}
 ds^{2}_{\mathcal{M}}=\frac{1-k^{2}}{\textrm{cn}^{2}\left(\sigma|k^{2}\right)}\left(d\sigma^{2}+d\tau^{2}\right)\,,
\qquad
-\mathbb{K}\left(k^{2}\right)<\sigma<\mathbb{K}\left(k^{2}\right)\,,
\qquad\tau\in\mathbb{R}\,,
\end{gather}
 where $\sigma, \tau$ are related to $\rho,t$ by  
\be\label{lines_AdS_relation}
\cosh \rho = \frac{\sqrt{1+b^2}}{b\, \textrm{cn}\left(\sigma|k^{2}\right)}\,,
\qquad\qquad
t=\frac{b\,p}{\sqrt{b^{4}+p^{2}}}\, \tau\,.
\ee
Introducing large cutoffs $ 0 \leq  \rho \leq \rho_0,\  \ 0 < t \leq T$ 
translates into 
\be\label{cutoff_tau_sigma}
\sigma\in(-\sigma_0,\,\sigma_0)\,,~ \quad 
\tau\in[0,\mathcal{T}] \ , \quad \qquad 
\sigma_0\equiv \textrm{cn}^{-1}\Big(\frac{\sqrt{1+b^2}}{b \cosh\rho_0}|k^2\Big) \,,~ \quad 
\mathcal{T} \equiv \frac{\sqrt{b^{4}+p^{2}}}{b\,p} \, T \ . 
\ee
 The  classical  string  action  (the   first term in  \eqref{3.1}) proportional  to the 
  regularized area of the surface 
 is   given, after    the subtraction of the divergence due to the two boundary lines at $\rho=\rho_0\to \infty$, 
  in terms of elliptic integrals~\cite{Drukker:2011za}
\begin{flalign}\no 
\Gamma^{(0)}(k)
&
= \frac{{1}}{2\pi} \int_{0}^{\mathcal{T}} d\tau\int_{-\sigma_0}^{\sigma_0}d\sigma
\frac{1-k^{2}}{\textrm{cn}^{2}(\sigma|k^{2})}\\
&\te 
= \frac{T}{2\pi} \Big[e^{\rho_0}+\frac{2\sqrt{b^4+p^2}}{b\,p} \left(\frac{(b^2+1)p^2}{b^4+p^2} \mathbb{K}(k^2)-\mathbb{E}(k^2) \right)+O(e^{-\rho_0})\Big]\no\\
&\  \to \frac{T}{2\pi} \frac{2\sqrt{b^4+p^2}}{b\,p} \Big[\frac{(b^2+1)p^2}{b^4+p^2} \mathbb{K}(k^2)-\mathbb{E}(k^2) \Big]\,.
\label{3.56}
\end{flalign}
The one-loop  effective action reads formally (cf. \rf{3.9})~\cite{Forini:2010ek,Drukker:2011za}
\begin{gather}\label{3.57}
\Gamma^{(1)}\left(k\right)=-\log \frac{\textrm{Det}^{8/4}[\mathcal{O}_{F}^{2}(k)]}{\textrm{Det}^{5/2}[\mathcal{O}_{0}(k)]\, \textrm{Det}^{2/2}[\mathcal{O}_{1}(k)] \,\textrm{Det}^{1/2}[\mathcal{O}_{2}(k)]}
\end{gather}
with the bosonic and the fermionic fluctuation operators given by 
\begin{flalign}
\label{libo_O0}
\mathcal{O}_{0}\left(k\right) & \equiv\frac{\textrm{cn}^{2}\left(\sigma|k^2\right)}{1-k^{2}}\left(-\partial_{\sigma}^{2}-\partial_{\tau}^{2}\right)\,, \qquad \qquad
\mathcal{O}_{1}\left(k\right)  \equiv\mathcal{O}_{0}\left(k\right)+2\,,\\
\label{libo_O2}
\mathcal{O}_{2}\left(k\right) & \equiv\mathcal{O}_{0}\left(k\right)+2-2\frac{k^{2}\textrm{cn}^{4}\left(\sigma|k^2\right)}{1-k^{2}}\,,\\
\label{lifo}
\mathcal{O}_{F}\left(k\right) & \equiv-i\frac{\textrm{cn}\left(\sigma|k^{2}\right)}{\sqrt{1-k^{2}}}\sigma_{1}\Big(\partial_{\sigma}+\frac{\textrm{sn}\left(\sigma|k^{2}\right)\textrm{dn}\left(\sigma|k^{2}\right)}{2\textrm{cn}\left(\sigma|k^{2}\right)}\Big)
 -i \frac{\textrm{cn}\left(\sigma|k^{2}\right)}{\sqrt{1-k^{2}}}\sigma_{2}\partial_{\tau}+\sigma_{3}\,.
\end{flalign}
The limiting case  of $k=0$  ($\phi=0$)  corresponds to a surface $\bar{\mathcal{M}}$ stretching
 between a pair of lines that are antipodal in $\mathbb{R}\times S^3$ \foot{Considering the theory in $\mathbb{R}^4$,
  related to the theory in $\mathbb{R}\times S^3$ by the stereographic projection, this is the infinite  straight line.}
   at the $AdS$ boundary, a configuration for which the corresponding Wilson loop is a  1/2  BPS protected observable with 
   the expectation value equal to one~\cite{Zarembo:2002an}.
Thus   the natural   choice for the expansion parameter $\a$ is
\be\la{3.60}
\alpha_{_{\rm cusp}}\equiv k^2\,.
\ee
In this case  the world-sheet cutoffs in \eqref{cutoff_tau_sigma}  also depend  on $\a$ and that  may be confusing. 
A sensible expansion would require  introducing new world-sheet coordinates $r,w$ with the  range independent of $k$. 
For the world-sheet time, one can simply  choose it  to  be  the $AdS$ time
\begin{gather}\label{w}
w\equiv t=\frac{b\, p}{\sqrt{b^{4}+p^{2}}}\,\tau\,,\qquad 
\qquad
w\in[0,\,T]\,,
\end{gather}
while finding a suitable spatial world-sheet coordinate appears to be more problematic~\footnote{For instance, we discard $\rho$ because its minimum value $\textrm{arccosh}({\sqrt{1+b^2}}/{b})$ is a function of $k$, and  the relation \eqref{lines_AdS_relation} between $\sigma$ and $\rho$ is not one-to-one. 
Another possibility is $r'\equiv{\pi}\sigma/({2 \sigma_0})$ which varies in the constant interval $(-{\pi}/{2},{\pi}/{2})$, however this choice would introduce the cutoff $\rho_0$ via $\sigma_0$ into \eqref{lines_metric} and \eqref{libo_O0}-\eqref{lifo} once the change of coordinates is made. This implies that the metric at $k=0$ is still dependent on one parameter and cannot have the geometry of $H^2$. The perturbative analysis for small $k$ would be then problematic, as the procedure relies on the knowledge of the heat kernels at $k=0$, which in this case one would still need to evaluate.}.
A good candidate~\cite{Aguilera-Damia:2014bqa} in the~$\rho_0\to\infty$  limit is
\begin{gather}\label{r}
r\equiv \frac{\pi \sigma}{2 \mathbb{K}(k^2)}\,,
\qquad\qquad 
r\in \left(-\frac{\pi}{2},\, \frac{\pi}{2}\right)\,,
\end{gather}
as in this limit $\sigma_0=\mathbb{K}(k^2)$, see \eqref{cutoff_tau_sigma}. At finite and large $\rho_0$, however,  the maximum value of $|r|$ is ${\pi \sigma_0}/({2 \mathbb{K}(k^2)})={\pi}/{2}+O(e^{-\rho_0})$ and $k$ reappears in the exponentially suppressed terms. 
We will later take into account that the integrals over $r$ may generate such $k$-dependent contributions (see footnote \ref{footnote_range}). 

In the limiting case $k=0$    eqs.  \eqref{lines_AdS_relation} and \eqref{w}-\eqref{r} simplify to $\sinh\rho=|\tan\sigma |=|\tan r |$ and $t=\tau=w$, the cutoffs \eqref{cutoff_tau_sigma} become $\sigma_0=\textrm{arctan}(\sinh\rho_{0})$ and $\mathcal{T} = T$, and \eqref{lines_metric} reduces to that of  the infinite-strip parametrization  or $H^2$   that we will call $\hat H^2$   
(with boundary $\mathbb{R}$  instead of $S^1$)
\begin{gather}
ds^{2}_{\bar{\mathcal{M}}}=\frac{1}{\cos^{2}r} (dr^{2}+dw^{2})\,. \la{3.63} 
\end{gather}
In this case the regularized volume  or the     value of string action vanishes (cf. \eqref{3.56})
\begin{gather}\label{volume_lines}
\Gamma^{(0)} (0) = { 1 \ov 2 \pi} V_{\hat H^{2}}= { 1 \ov 2 \pi} 
\int_{0}^{T}dw\int_{-\textrm{arctan}(\sinh\rho_{0})}^{\textrm{arctan}(\sinh\rho_{0})}\frac{dr}{\cos^{2}r}={T\ov 2\pi} \big[ 
e^{\rho_{0}}+O(e^{-\rho_0})\big]  \ \to\ 0\, , 
\end{gather}
  in agreement with the $k=0$ limit of  \eqref{3.56}.
For $k=0$ the operators \eqref{libo_O0}-\eqref{lifo} become those of the straight line  Wilson loop 
~\cite{Forste:1999qn,Drukker:2000ep} 
\begin{gather}\label{liop_0}
\bar{\mathcal{O}}_{0}  =-\Delta_{r,w}\,,\qquad\qquad 
\bar{\mathcal{O}}_{1}=\bar{\mathcal{O}}_{2}  =-\Delta_{r,w}+2\,,\qquad\qquad 
\bar{\mathcal{O}}_{F}  =-i\slashed{\nabla}_{r,w}+\sigma_{3}\ , 
\end{gather}
with the Laplacian given in \eqref{strip_Laplace_operator} and  the Dirac operator in \eqref{strip_Dirac_operator}. 
Here the  multiplicities  and the masses coincide with those in the spectrum \eqref{la_operator_order_0}  corresponding to a 
 circular Wilson loop in $\mathbb{R}^4$. 
 The zeta-functions \eqref{ze_scalar} of these operators are proportional to the  volume $V_{\hat H^2}$  whose 
 renormalized   value is  zero \eqref{volume_lines} and thus we get  \cite{Drukker:2000rr,Buchbinder:2014nia}~\footnote{The minimal surface \eqref{3.5}-\eqref{3.6} bounded by a circle and the one ending 
 on a straigh line or  two antiparallel  lines have the same  local geometry 
 of $H^2$, as they 
 are mapped to each other through an isometry of $AdS_5$. 
 The difference in the values of their regularized 
 volumes \eqref{3.17} and \eqref{volume_lines}  is a regularization effect due to the
  different  global properties of the two spaces -- different topology of the boundary
  (see  \cite{Drukker:2000rr,Buchbinder:2014nia,Bergamin:2015vxa}   for a discussion of this point).}
 \begin{flalign}\label{3.57_order_0}
{\Gamma}^{(1)}(0) \te 
=-\frac{5}{2}\bar{\zeta}^{'}_{\mathcal{O}_{0}}(0)-\frac{2}{2}\bar{\zeta}^{'}_{\mathcal{O}_{1}}(0)-\frac{1}{2}\bar{\zeta}^{'}_{\mathcal{O}_{2}}(0)+\frac{8}{4}\bar{\zeta}^{'}_{\mathcal{O}_{F}^{2}}(0) =0\,.
\end{flalign}
 For small values of $\alpha_{_{\rm cusp}}=k^2$ the angle  in  \eqref{lines_phi}   is also small, 
      $\phi=\pi k +O(k^3)$. 
      In this near-BPS limit we  get, expanding  the elliptic integral  in the metric \eqref{lines_metric} (cf. \rf{expansion})
\begin{flalign}
\label{lines_metric_expansion}
\bar{g}_{ij}(r,w)&=
\frac{1}{\cos^{2}r}
\left(\begin{array}{cc}
1 & 0\\
0 & 1
\end{array}\right)\,,
\qquad
\tilde{g}_{ij}(r,w)=
\left(\begin{array}{cc}
-\frac{1}{2} & 0\\
0 & \frac{3}{2\cos^{2}r}-\frac{1}{2}
\end{array}\right)\,.
\end{flalign}
The   order $k^2$ terms  in the operators  \eqref{libo_O0}--\eqref{lifo}   are found to be 
barred operators given by \eqref{liop_0} and their perturbations by
\begin{flalign}
\label{libo_O0_order_1}
\tilde{\mathcal{O}}_{0}& =\tilde{\mathcal{O}}_{1} \te  =\frac{\cos^{2}r}{2}\Big[-\cos^{2}r\partial_{r}^{2}+\left(2+\sin^{2}r\right)\partial_{w}^{2}\Big]\,,\\
\label{libo_O2_order_1}
\tilde{\mathcal{O}}_{2} &\te  =\frac{\cos^{2}r}{2}\Big[-\cos^{2}r\partial_{r}^{2}+\left(2+\sin^{2}r\right)\partial_{w}^{2}-4\cos^{2}r\Big]\,,\\
\label{lifo_order_1}
\tilde{\mathcal{O}}_{F} &\te  =-\frac{i\cos^{3}r}{4}\sigma_{1}\partial_{r}-\frac{i\left(\cos3r-9\cos r\right)}{16}\sigma_{2}\partial_{w}-\frac{3i\sin r\cos^{2}r}{8}\sigma_{1}\,.
\end{flalign}
As in \rf{lbf_squared_expansion},  we will  actually  be using the expansion of   the   square  of  the fermionic operator: 
\be \mathcal{O}_{F}^{2}(k)  =\bar{\mathcal{O}}_{F}^{2}+k^{2}\,\{ \bar{\mathcal{O}}_{F},\tilde{\mathcal{O}}_{F}\}+O(k^4) \ . \la{3.71} 
\ee 
%
For each operator in \eqref{3.57} we will repeat a procedure similar to that explained 
between  \eqref{fe_1}--\eqref{ze_O1_order_1}, with two differences. Since we rescaled the world-sheet coordinates \eqref{w}--\eqref{r} differently, none of the operators \eqref{libo_O0_order_1}--\eqref{lifo_order_1} can be written in terms of the Laplacian \eqref{strip_Laplace_operator} or the Dirac operator \eqref{strip_Dirac_operator}. 
Therefore we will use the full heat kernels \eqref{B.11} and \eqref{b13} instead of their simpler expressions at coincident points. Also, in the integrals over the (regularized) world-sheet, the domain of integration of $r$ depends on the perturbative parameter $k$, and  divergences appear if the radial cutoff $\rho_0\to\infty$ is removed at fixed $k$. 
By analogy with  \eqref{volume_lines}, we shall assume that a 
sensible regularization at small $k$ consists in doing  the integrals for finite $\rho_0$, expanding  in 
 $\rho_0\to\infty$ and dropping  all positive powers of $e^{\rho_0}$. 
  It is easy to check that  since negative powers of $k^2$ are absent,
   in what is left we   can simply take the limit $k\to 0$ 
    (see also footnote \ref{footnote_range}). 
Applying this to the bosonic operator $\mathcal{O}_0=  \mathcal{\bar O}_0  + k^2  \mathcal{\tilde O}_0  + ...   $  we find for the correction to its heat kernel 
 (see \rf{2.8}) 
\begin{flalign}\no 
\tilde{K}_{\mathcal{O}_{0}}(t) & =-{t} \int_{-\frac{\pi \sigma_0}{2 \mathbb{K}(k^2)}}^{\frac{\pi \sigma_0}{2 \mathbb{K}(k^2)}} dr \int_0^{T} 
\frac{dw}{\cos^{2}r}\Big[\tilde{\mathcal{O}}_{0}\bar{K}_{-\Delta}(r,w,r',w';t)\Big]_{r=r',w=w'}\\
 & ={\te \frac{1}{8\pi}}  \Big[\big(3e^{\rho_{0}}-2\pi+O(e^{-\rho_{0}})\big)+O(k^2)\Big]t\,T\int_{0}^{\infty}dv\,v\tanh\pi v\  {\te (v^{2}+\frac{1}{4})} \ e^{-t\big(v^{2}+\frac{1}{4}\big)}\nonumber\\
 & \to-\frac{t\,T}{4}\int_{0}^{\infty}dv\,v\tanh\pi v  \  {\te (v^{2}+\frac{1}{4})} \  e^{-t(v^{2}+\frac{1}{4})}\,,  \label{fe_8}
\end{flalign}
where we used that ${\pi \sigma_0}/({2 \mathbb{K}(k^2)})=\textrm{arctan}(\sinh\rho_{0})+O(k^2)$ after taking the large-$\rho_0$ limit.
The corresponding zeta-function is
\begin{flalign}\no 
\tilde{\zeta}_{\mathcal{O}_{0}}(s) 
 & =-\frac{s\,T}{4}\int_{0}^{\infty}dv\frac{1}{\big(v^{2}+\frac{1}{4}\big)^{s}}\Big(v-\frac{2v}{e^{2\pi v}+1}\Big) \\
 & =-\frac{s\,T}{8\big(s-1\big)}\big(\frac{1}{4}\big)^{1-s}+\frac{s\,T}{2}\int_{0}^{\infty}dv\frac{v}{\big(e^{2\pi v}+1\big)\big(v^{2}+\frac{1}{4}\big)^{s}}\label{fe_12} \ ,  \\
\tilde{\zeta}_{\mathcal{O}_{0}}^{'}(0) & ={\te \frac{1}{24} } T ~.
\end{flalign}
Similarly, for the remaining bosonic and fermionic operators one gets 
\begingroup
\allowdisplaybreaks
\begin{flalign}\label{fe_20}
\tilde{K}_{\mathcal{O}_{1}}(t) & =-{t}\int_{-\frac{\pi \sigma_0}{2 \mathbb{K}(k^2)}}^{\frac{\pi \sigma_0}{2 \mathbb{K}(k^2)}} dr \int_0^{T} 
\frac{dw}{\cos^{2}r}\Big[\tilde{\mathcal{O}}_{1}\bar{K}_{-\Delta+2}(r,w,r',w';t)\Big]_{r=r',w=w'}\  \\
 & =-\frac{t\,T}{4}\int_{0}^{\infty}dv\,v\tanh\pi v\ \big(v^{2}+{\te \frac{1}{4}\big)} \ e^{-t(v^{2}+\frac{9}{4})}\ , \nonumber \\
\label{fe_13}
\tilde{\zeta}_{\mathcal{O}_{1}}(s) 
 & =-\frac{s\,T}{4}\int_{0}^{\infty}dv\frac{1}{\big(v^{2}+\frac{9}{4}\big)^{s}}\big(1-\frac{2}{v^{2}+\frac{9}{4}}\big)\big(v-\frac{2v}{e^{2\pi v}+1}\big) \\
 & =-\frac{s\,T}{8\big(s-1\big)}\big(\frac{9}{4}\big)^{1-s}+\frac{T}{4}\big(\frac{9}{4}\big)^{-s}
+\frac{s\,T}{2}\int_{0}^{\infty}dv\frac{v}{\big(e^{2\pi v}+1\big)\big(v^{2}+\frac{9}{4}\big)^{s}}\no\\
&\quad -{s\,T}\int_{0}^{\infty}dv\frac{v}{\big(e^{2\pi v}+1\big)\big(v^{2}+\frac{9}{4}\big)^{s+1}}  
\label{zetaprimeO1} \ , \qquad 
\tilde{\zeta}_{\mathcal{O}_{1}}^{'}(0) 
  =\te  \big( -{5\over 24} +{ 1 \ov 2} \gamma\big)  T \,,\\
\label{fe_21}
 \tilde{K}_{\mathcal{O}_{2}}(t) & =-{t}
\int_{-\frac{\pi \sigma_0}{2 \mathbb{K}(k^2)}}^{\frac{\pi \sigma_0}{2 \mathbb{K}(k^2)}} dr \int_0^{T} 
\frac{dw}{\cos^{2}r}\Big[\tilde{\mathcal{O}}_{2}\bar{K}_{-\Delta+2}(r,w,r',w';t)\Big]_{r=r',w=w'}\\
 & =\frac{t\,T}{4}\int_{0}^{\infty}dv\,v\tanh\pi v\Big[-\big(v^{2}+\frac{1}{4}\big)+2\Big]e^{-t\big(v^{2}+\frac{9}{4}\big)}\ , \nonumber \\
\tilde{\zeta}_{\mathcal{O}_{2}}\big(s\big) 
 & =\frac{s\,T}{4}\int_{0}^{\infty}dv\frac{1}{\big(v^{2}+\frac{9}{4}\big)^{s}}\big(-1+\frac{4}{v^{2}+\frac{9}{4}}\big)\big(v-\frac{2v}{e^{2\pi v}+1}\big) \\
 & =-\frac{s\,T}{8\big(s-1\big)}\big(\frac{9}{4}\big)^{1-s}+\frac{T}{2}\big(\frac{9}{4}\big)^{-s}
  +\frac{s\,T}{2}\int_{0}^{\infty}dv\frac{v}{\big(e^{2\pi v}+1\big)\big(v^{2}+\frac{9}{4}\big)^{s}}\no\\
& \quad  -{2 s\,T}\int_{0}^{\infty}dv\frac{v}{\big(e^{2\pi v}+1\big)\big(v^{2}+\frac{9}{4}\big)^{s+1}}
\ , \qquad \label{zetaprimeO2}
\tilde{\zeta}_{\mathcal{O}_{2}}^{'}(0) 
  =\te \big( - \frac{17}{24} +\gamma \big)  T\ ,  \\
\tilde{K}_{\mathcal{O}_{F}}(t) & =-t
\int_{-\frac{\pi \sigma_0}{2 \mathbb{K}(k^2)}}^{\frac{\pi \sigma_0}{2 \mathbb{K}(k^2)}} dr \int_0^{T} 
\frac{dw}{\cos^{2}r}\textrm{tr}\Big[\left\{ \bar{\mathcal{O}}_{F}^{r,w},\tilde{\mathcal{O}}_{F}^{r,w}\right\} \bar{K}_{-\slashed{\nabla}^{2}+1}(r,w,r',w';t)\Big]_{r=r',w=w'}\   \no\\
\label{fe_22}
& =\frac{t\,T}{2}\int_{0}^{\infty}dv\,v\coth\pi v\Big[1-\big(v^{2}+1\big)\Big]e^{-t\big(v^{2}+1\big)}\ , \\
\label{fe_14}
\tilde{\zeta}_{\mathcal{O}_{F}^{2}}(s)
 & =\frac{s\,T}{2}\int_{0}^{\infty}dv\frac{1}{\big(v^{2}+1\big)^{s}}\big(\frac{1}{v^{2}+1}-1\big)\big(v+\frac{2v}{e^{2\pi v}-1}\big) \\
 & =-\frac{s\,T}{4\big(s-1\big)}+\frac{T}{4}
-\int_{0}^{\infty}dv\frac{{s\,T} v}{\big(e^{2\pi v}-1\big)\big(v^{2}+1\big)^{s}}  +\int_{0}^{\infty}dv\frac{{s\,T} v}{\big(e^{2\pi v}-1\big)\big(v^{2}+1\big)^{s+1}}\nonumber ,  \\\label{zetaprimeOF}
\tilde{\zeta}_{\mathcal{O}_{F}^{2}}^{'}(0) 
 & =\te \big(-\frac{1}{24}+\frac{\gamma}{2}\big)T \ , 
\end{flalign}%
\endgroup
where \eqref{formula6} was used  to compute \eqref{zetaprimeO1} and \eqref{zetaprimeO2}, and \eqref{formula8}   -- to find 
 \eqref{zetaprimeOF}. 
 
The resulting one-loop effective action is then 
\begin{flalign}\label{Delta_lines_order_1}
\Gamma^{(1)}\left(k\right)- \Gamma^{(1)}\left( 0\right)&=\te 
{k^2}\left( -\frac{5}{2}\tilde{\zeta}^{'}_{\mathcal{O}_{0}}(0)-\frac{2}{2}\tilde{\zeta}^{'}_{\mathcal{O}_{1}}(0)-\frac{1}{2}\tilde{\zeta}^{'}_{\mathcal{O}_{2}}(0)+\frac{8}{4}\tilde{\zeta}^{'}_{\mathcal{O}_{F}^{2}}(0)\right)+O(k^4)\no\\
&\te  =\frac{3}{8} T k^2 +O(k^4)\equiv \frac{3}{8\,\pi^2} T \phi^2 +O(\phi^4)\ , \end{flalign} 
where in the last line we substituted the expansion of~\eqref{lines_phi}. 
This reproduces, as it should,  the result  of~\cite{Drukker:2011za} for the so-called Bremsstrahlung function~\cite{Correa:2012at}.

As in the case of the latitude Wilson loop,  
it is not difficult to check that considering the sum of perturbed contributions to the zeta-functions  
\be\label{la_zf_ac_(3)}  \tilde{\zeta}_{\rm tot} (s) = \te 
\frac{5}{2}\tilde{\zeta}_{\mathcal{O}_{0}}(s)+\frac{2}{2}\tilde{\zeta}_{\mathcal{O}_{1}}(s)+\frac{1}{2}\tilde{\zeta}_{\mathcal{O}_{2}}(s)-\frac{8}{4}\tilde{\zeta}_{\mathcal{O}_{F}^{2}}(s)
\ee
eliminates the need of an analytical continuation in $s$: setting $s=0$ in the total integrand   and then performing the 
 integration gives  \eqref{Delta_lines_order_1}. 
 This is again consistent with the fact that the trace of the full heat kernel, which equals to  the sum of \eqref{fe_8}, \eqref{fe_20}, \eqref{fe_21} and \eqref{fe_22},  vanishes  for small $t$
\begin{gather}\label{fe_31}
\tilde{K}_{\rm tot} (t) = {\te 
\frac{5}{2}\tilde{K}_{\mathcal{O}_{0}}(t)+\frac{2}{2}\tilde{K}_{\mathcal{O}_{1}}(t)+\frac{1}{2}\tilde{K}_{\mathcal{O}_{2}}(t)-\frac{8}{4}\tilde{K}_{\mathcal{O}_{F}}(t)}=\frac{0}{t}+0+O(t)\ ,
\end{gather}
which implies that  \eqref{la_zf_ac_(3)} does not develop any singularity in $s=0$.

\subsection{$k$-wound circular Wilson loop}
\label{sec:kcircle}

Our next example is 
the  minimal surface   generalizing  the circular Wilson loop   one (given by the $\theta_0=0$ limit of
 \eqref{3.5}-\eqref{3.6}) to the case  of  an arbitrary integer 
  winding number $k$ 
  along the circle.  
   The   string theory solution should be representing, at strong coupling, the gauge-theory 
      circular Wilson loop in the $k$-fundamental representation.

       This   classical  solution   can be  found simply by 
   the replacements $\sigma\to k \sigma$ and $\tau \to k \tau$ in  \rf{3.15}  ~\cite{Drukker:2005cu,Kruczenski:2008zk}, 
so that  the induced metric  becomes 
\begin{gather}\label{kcircle_metric}
ds^{2}
=\Omega^2(\sigma) \left(d\sigma^{2}+d\tau^{2}\right)\,, \qquad \Omega(\sigma) = \frac{k}{\sinh(k \sigma)} \ , \qquad 
\qquad \sigma\in[0,\infty)\,,
\qquad \tau\in [0,2\pi)\,.
\end{gather}
%
The  corresponding geometry is a cone of $AdS_2$  with negative angular deficit $\delta=2\pi(1-k)$. 
Given a singular nature of this geometry  one   may wonder if  a  
perturbation theory near   $k=1$  limit is   meaningful. 
We will  first proceed   formally and  then comment on  possible   issues   at the end of this section.

The relation $z=\tanh(k\sigma)$ from \eqref{3.5} implies that the world-sheet coordinate $\sigma$ is to be cut off at $k^{-1}\arctanh\,\epsilon$ in order keep the same physical cutoff at $z=\epsilon$ for any value of $k$. 
Then the classical string   action is  
~\cite{Kruczenski:2008zk}  (cf. \rf{3.17})
\begin{gather}\label{3.88}
\Gamma^{(0)}(k)
= \frac{1}{2\pi} \int_{0}^{2\pi} d\tau\int_{k^{-1} \arctanh\, \epsilon}^{\infty}d\sigma \,\, \frac{k^2}{\sinh^2(k \sigma)}
=\frac{k}{\epsilon}-k \ 
\to-k \,.
\end{gather}
One  may  define   the new coordinate $\rho$   that ranges in the  same   interval $[\textrm{arccosh}({\epsilon}^{-1}),\infty)$ for any $k$
 by 
\be\la{3.92} 
\sinh\rho\equiv({\sinh(k\sigma)})^{-1} \ . 
\ee 
The  one-loop correction in \rf{3.1}  is ~\cite{Kruczenski:2008zk,Bergamin:2015vxa} 
\begin{gather}\label{kcircle_one_loop_eff_action}
\Gamma^{(1)}(k)=-\log\frac{\Det^{5/2} [\mathcal{O}_{0}(k)]\, \Det^{3/2} [\mathcal{O}_{1}(k)]}{\Det^{8/4} [\mathcal{O}_{F}(k)]}\ ,
\end{gather}
where 
\begin{flalign}\label{kbo}
\mathcal{O}_{0}\left(k\right) & \equiv\frac{\sinh^{2}(k\sigma)}{k^{2}}\left(-\partial_{\tau}^{2}-\partial_{\sigma}^{2}\right)\,,
\qquad\qquad
\mathcal{O}_{1}\left(k\right) \equiv\mathcal{O}_{0}\left(k\right)+2\,,\\
\label{kfo}
\mathcal{O}_{F}\left(k\right) &\te  \equiv \frac{i}{k} \sinh(k\sigma)\sigma_{1}\big[ \partial_{\sigma}-\frac{k}{2} \coth(k\sigma)\big]
 -\frac{i}{k} \sinh(k\sigma)\, \sigma_{2}\partial_{\tau}+\sigma_{3}\,. 
\end{flalign}
For $k=1$ the corresponding world-sheet surface \eqref{kcircle_metric} becomes  that of  $\bar{\mathcal{M}}=H^2$,  i.e. \eqref{3.15},  
with the boundary  $S^1$ at  $\rho=\textrm{arccosh}({\epsilon}^{-1})$ and the regularized area in \eqref{3.17}.
 The spectrum of  excitations then coincides with \eqref{la_operator_order_0}
\begin{gather}\label{kcircle_operator_order_0}
\bar{\mathcal{O}}_{0}  =-\Delta_{\rho,\tau}\,, \qquad
\bar{\mathcal{O}}_{1}  =-\Delta_{\rho,\tau}+2\,, \qquad
\bar{\mathcal{O}}_{F}  =-i\slashed{\nabla}_{\rho,\tau}+\sigma_{3} 
\end{gather}
so that  the  1-loop   correction in \rf{3.1}  is also the same  as in \eqref{3.9_order_0}
\begin{flalign}\label{kcircle_one_loop_eff_action_order_0}
\Gamma^{(1)}\left(k=1\right)\te = -\frac{5}{2}\bar{\zeta}^{'}_{\mathcal{O}_{0}}(0)-\frac{3}{2}\bar{\zeta}^{'}_{\mathcal{O}_{1}}(0)+\frac{8}{4}\bar{\zeta}^{'}_{\mathcal{O}_{F}^{2}}(0)
= \frac{1}{2}\log2\pi~.
\end{flalign}
%
For $k=2,3,...$ the space \eqref{kcircle_metric} is a cone of $H^2$ with a 
conical singularity at $\rho=0$.  
We shall  formally   treat    $k$  as a real number  and expand in $k-1$,   i.e.  define the small  parameter $\a$ as 
\be\la{3.95}
\alpha_{_{\rm k-circle}}\equiv k-1\,.
\ee
%
 The small-$\a$  expansion of  the metric  \eqref{kcircle_metric} yields the leading and subleading terms as 
\begin{flalign}
\bar{g}_{ij}(\rho,\tau)&=
\left(\begin{array}{cc}
1 & 0\\
0 & \sinh^{2}\rho
\end{array}\right)\,,
~~~~\qquad 
\tilde{g}_{ij}(\rho,\tau)=
\left(\begin{array}{cc}
0 & 0\\
0 & 2\sinh^2\rho
\end{array}\right)\,.  
\end{flalign}
For the  leading-order  corrections  in the operators  \rf{kbo},\rf{kfo}   we find 
\begin{gather}\label{3.97}
\tilde{\mathcal{O}}_{0}=\tilde{\mathcal{O}}_{1} =\frac{2}{\sinh^2\rho} \partial_\tau^2\,,
\qquad\qquad
\tilde{\mathcal{O}}_{F}=\frac{i}{\sinh\rho} \sigma_2 \partial_\tau\,.
\end{gather}
%
In the perturbative expansion in $k-1$ of the heat kernels and zeta-functions 
%
 the  integrals  will contain similar  $1\ov \epsilon$ divergences  as in the  volume \rf{3.88}.
 As  in 
 Section \ref{sec:generalized_cusp},   we will  first 
  compute  the integrals at finite cutoff,  then take the limit $\epsilon\to 0$ in the result and finally 
   drop terms with negative  powers of $\epsilon$.
   Using this regularization prescription 
    we find  (here $\Lambda= \textrm{arccosh}({\epsilon}^{-1})$ as in \rf{3.155}) 
 \begingroup
\allowdisplaybreaks
\begin{flalign}\no 
\tilde{K}_{\mathcal{O}_{0}}\left(t\right) & =-t \int_{0}^\Lambda\, d\rho \int_{0}^{2\pi}d\tau \frac{2}{\sinh\rho} \Big[\partial_\tau^2 \bar{K}_{-\Delta}(\rho,\tau,\rho',\tau';t)\Big]_{\rho=\rho',\tau=\tau'}\\
 &  =-{t}\int_{0}^{\infty}dv\,v\tanh\left(\pi v\right){ \te \left(v^{2}+\frac{1}{4}\right) \ }e^{-t\left(v^{2}+\frac{1}{4}\right)}\ , 
\label{fe_28}\\ \no 
\tilde{\zeta}_{\mathcal{O}_{0}}\left(s\right)
 &
\te  = \int_{0}^{\infty}dv\frac{-sv}{\left(v^{2}+\frac{1}{4}\right)^{s}}+\int_{0}^{\infty}dv\frac{2sv}{\left(e^{2\pi v}+1\right)\left(v^{2}+\frac{1}{4}\right)^{s}}\\
 &\te  =-\frac{s}{2\left(s-1\right)}\left(\frac{1}{4}\right)^{1-s}+2s\int_{0}^{\infty}dv\frac{v}{\left(e^{2\pi v}+1\right)\left(v^{2}+\frac{1}{4}\right)^{s}}\label{fe_15} \ , \qquad 
\tilde{\zeta}_{\mathcal{O}_{0}}^{'}\left(0\right)
  =\frac{1}{6} \ , \\
\label{fe_29}
\tilde{K}_{\mathcal{O}_{1}}\left(t\right) &=-t \int_{0}^\Lambda\, d\rho \int_{0}^{2\pi}d\tau \frac{2}{\sinh\rho} \Big[\partial_\tau^2 \bar{K}_{-\Delta+2}(\rho,\tau,\rho',\tau';t)\Big]_{\rho=\rho',\tau=\tau'}\no\\
 &  =-{t}\int_{0}^{\infty}dv\,v\tanh\left(\pi v\right) {\te \left(v^{2}+\frac{1}{4}\right)} \ e^{-t\left(v^{2}+\frac{9}{4}\right)} \ , \\ \no 
  \tilde{\zeta}_{\mathcal{O}_{1}}\left(s\right) 
 & 
= \textstyle  \int_{0}^{\infty}dv\frac{-sv}{\left(v^{2}+\frac{9}{4}\right)^{s}}\left(1-\frac{2}{v^{2}+\frac{9}{4}}\right)+\int_{0}^{\infty}dv\frac{2sv}{\left(e^{2\pi v}+1\right)\left(v^{2}+\frac{9}{4}\right)^{s}}\left(1-\frac{2}{v^{2}+\frac{9}{4}}\right)\\
 &\te  =-\frac{s}{2\left(s-1\right)}\left(\frac{9}{4}\right)^{1-s}+\left(\frac{9}{4}\right)^{-s}
+2s\int_{0}^{\infty}dv\frac{v}{\left(e^{2\pi v}+1\right)\left(v^{2}+\frac{9}{4}\right)^{s}} \nonumber \\
 &\quad \te  -4s\int_{0}^{\infty}dv\frac{v}{\left(e^{2\pi v}+1\right)\left(v^{2}+\frac{9}{4}\right)^{s+1}} \ , \qquad \qquad 
\tilde{\zeta}_{\mathcal{O}_{1}}^{'}\left(0\right) 
 \te  =-\frac{5}{6}+2\gamma\ , \\
%
\tilde{K}_{\mathcal{O}_{F}^{2}}\left(t\right) & =- \int_{0}^\Lambda\, d\rho \int_{0}^{2\pi}d\tau\sinh\rho\, \textrm{tr}\Big[\left\{ \bar{\mathcal{O}}_{F}^{\rho,\tau},\tilde{\mathcal{O}}_{F}^{\rho,\tau}\right\} \bar{K}_{-\slashed{\nabla}^{2}+1}(\rho,\tau,\rho',\tau';t)\Big]_{\rho=\rho',\tau=\tau'}\no\\
\label{fe_30}
 & =-t\int_{0}^{\infty}dv\,v\coth\left(\pi v\right)\left(2v^{2}+1\right)e^{-t\left(v^{2}+1\right)} \ , \\
\no 
\tilde{\zeta}_{\mathcal{O}_{F}^{2}}\left(s\right)
 & 
=  \textstyle \int_{0}^{\infty}dv\,\frac{-sv}{\left(v^{2}+1\right)^{s}}\left(2-\frac{1}{v^{2}+1}\right)+\int_{0}^{\infty}dv\,\frac{-2sv}{\left(e^{2\pi v}-1\right)\left(v^{2}+1\right)^{s}}\left(2-\frac{1}{v^{2}+1}\right)\\
 &\te  =-\frac{s}{s-1}+\frac{1}{2}-4s\int_{0}^{\infty}dv\,\frac{v}{\left(e^{2\pi v}-1\right)\left(v^{2}+1\right)^{s}}+2s\int_{0}^{\infty}dv\,\frac{v}{\left(e^{2\pi v}-1\right)\left(v^{2}+1\right)^{s+1}} \ ,  \\
\tilde{\zeta}^{'}_{\mathcal{O}_{F}^{2}}\left(s\right) 
 &\te  =\frac{1}{3}+\gamma  \ , 
\end{flalign}%
\endgroup
where  we used  \eqref{formula6} and \eqref{formula8}.
Combining these results, the one-loop effective action reads
\begin{flalign}\label{3.105}
&\Gamma^{(1)}\left(k\right)- \Gamma^{(1)}\left(k=1\right)=  c_1 (k-1)+O\big((k-1)^2\big) \ , \\
& c_1 = 
\te 
 -\frac{5}{2}\tilde{\zeta}^{'}_{\mathcal{O}_{0}}(0)-\frac{3}{2}\tilde{\zeta}^{'}_{\mathcal{O}_{1}}(0)+\frac{8}{4}\tilde{\zeta}^{'}_{\mathcal{O}_{F}^{2}}(0)= \frac{3}{2}-\gamma \ . \la{3.106} 
\end{flalign}
At the same time, the  strong-coupling expansion of the 
 gauge theory prediction for the  expectation value  $\langle\mathcal{W}\left(\lambda,k\right)\rangle = e^{-\Gamma(\l,k)}$  of    $k$-fundamental circular loop normalized to the $k=1$ value 
  is  (cf. \rf{1.1})~\cite{Drukker:2005cu,Pestun:2007rz}
\be\label{3.107}
\Gamma(\lambda, k)-\Gamma(\lambda, k=1)
=\sqrt{\lambda}\,(1 - k)+{\te \frac{3}{2}}\,\log k+\mathcal{O}(\lambda^{-{1}/{2}}) ~.
\ee
Our string  theory result  \eqref{3.106}  thus coincides with the $k\to  1$ expansion of  the $\log k$ term in \rf{3.107} 
just  up to  an extra $\gamma$ (the Euler-Mascheroni constant)  term in $c_1$.

 Our  value  for $c_1= \frac{3}{2}-\gamma \approx 0.923$ may be compared to the results of the 
 two previous  string theory computations  of 
 $\Gamma^{(1)}\left(k\right)$ in \cite{Kruczenski:2008zk} and in \ci{Bergamin:2015vxa}.
 The 1-loop correction   in  \cite{Kruczenski:2008zk}  
 was 
  \be \Gamma^{(1)}_{\rm KT}\left(k\right)=\te  {1 \ov 2} \ln ( 2 \pi ) + (2 k + {1\over 2} ) \ln k - \ln \Gamma(k+1)  \ , \la{3.1077} \ee
 so that $(c_1)_{\rm KT}= { 3\ov 2} + \gamma \approx 2.077$,  which, surprisingly, differs from \rf{3.106}  just  by the sign 
 of the  $\gamma$ term. 
  This suggests   that  the presence of this extra $\gamma$ term in  both   approaches 
   is  a  regularization artifact  (see also  below). 
 The result of   \cite{Bergamin:2015vxa}  was   given  by  
 \ba \la{3.108}
&& \Gamma^{(1)}_{\rm  BT}\left(k\right)
 ={\te  \frac{1}{2} k \log (2 \pi )}  + I(k) \ , \ \ \ \ \ \ \   \\
 &&
I(k) =  -{\te \frac{1}{4}}\int_0^\infty\te \frac{dy}{y\,\sinh y} \Big[ \te (5 e^{-y}+3 e^{-3 y})
   \big( \coth {y\ov k}  - k \coth y\big)+16 e^{-2 y} (\frac{1}{\sinh {y \ov k}}-\frac{k}{\sinh y}) \Big] \ , \no 
 \ea
 so that  $(c_1)_{\rm BT}=  {1 \ov 2} \ln ( 2 \pi )  + I'(1) \approx 0.9189 +  0.3161= 1.235$  which is  closer but 
 still  different from the gauge-theory prediction $c_1= 1.5$  in \rf{3.106}.


From a   technical point of view, the presence of the extra $\gamma$ term in \rf{3.106} 
  can be  traced back to  the  dependence on a  regularization  used to define the   $s=0$ limit in the zeta-functions, which, 
   in contrast to   the  examples  in the previous two  subsections,  does   not cancel out 
   in the  sum  of  leading-order corrections  to the  zeta-functions. 
Splitting  the power and exponential terms in the integrands  in \eqref{fe_28}--\rf{fe_30}  (cf.  \rf{3.466}),   i.e. 
$\tilde{\zeta}_{\mathcal{O}}\equiv\tilde{\zeta}_{\mathcal{O}}^{(\textrm{power})}(s)+\tilde{\zeta}_{\mathcal{O}}^{(\textrm{exp})}(s)$,  
we  find 
\begin{flalign}\label{fe_27}
 & \te \frac{5}{2}\tilde{\zeta}_{\mathcal{O}_{0}}^{(\textrm{power})}(s)+\frac{3}{2}\tilde{\zeta}_{\mathcal{O}_{1}}^{(\textrm{power})}(s)-\frac{8}{4}\tilde{\zeta}_{\mathcal{O}_{F}^{2}}^{(\textrm{power})}(s)\\
& \qquad =  \int_{0}^{\infty}dv\te \Big[-\frac{5sv}{s\left(v^{2}+\frac{1}{4}\right)^{s}}
-\frac{3sv}{2\left(v^{2}+\frac{9}{4}\right)^{s}}\left(1-\frac{2}{v^{2}+\frac{9}{4}}\right)
+\frac{2sv}{\left(v^{2}+1\right)^{s}}\left(2-\frac{1}{v^{2}+1}\right)\Big]\,, \no
\end{flalign}
which is divergent for $s\to 0$. 
Proceeding  without performing an analytical continuation in $s$  gives 
\begin{flalign}\label{fe_34}
& \te \frac{d}{ds}\left(\te -\frac{5}{2}\tilde{\zeta}_{\mathcal{O}_{0}}(s)-\frac{3}{2}\tilde{\zeta}_{\mathcal{O}_{1}}(s)+\frac{8}{4}\tilde{\zeta}_{\mathcal{O}_{F}^{2}}(s)\right)_{
s=0}\\
 &\quad =
\int_0^{\infty}dv \te \frac{2v (2v^2-3)}{4v^4+13v^2+9}
+\frac{d}{ds}
\left(- \frac{5}{2}\tilde{\zeta}_{\mathcal{O}_{0}}^{(\textrm{exp})}(s)-\frac{3}{2}\tilde{\zeta}_{\mathcal{O}_{1}}^{(\textrm{exp})}(s)+\frac{8}{4}\tilde{\zeta}_{\mathcal{O}_{F}^{2}}^{(\textrm{exp})}(s) \right)_{s=0}\,,\no
\end{flalign}
 where  the   integral  diverges logarithmically for  large $v$.  
 This reflects the  presence of   $t^0$    term in the small-$t$  expansion of the  leading $k-1$ correction to the heat kernel 
 (cf. \rf{fe_322},\rf{fe_31})
\begin{gather}\label{fe_33} 
\te \tilde{K}_{\rm tot }(t)=
\frac{5}{2}\tilde{K}_{\mathcal{O}_{0}}(t)+\frac{3}{2}\tilde{K}_{\mathcal{O}_{1}}(t)-\frac{8}{4}\tilde{K}_{\mathcal{O}_{F}^{2}}(t)=\frac{0}{t}+\frac{1}{2}+O(t)
\end{gather}
where we used  that according to \eqref{fe_28}, \eqref{fe_29}, \eqref{fe_30}, 
\begin{gather}\label{fe_32}
\te \tilde{K}_{\mathcal{O}_{0}}(t) =-\frac{1}{2t}+O(t)\,,
~~~~ 
\tilde{K}_{\mathcal{O}_{1}}(t) =-\frac{1}{2t}+1+O(t)\,,
~~~~
\tilde{K}_{\mathcal{O}_{F}^{2}}(t) =-\frac{1}{t}+\frac{1}{2}+O(t)\,.
\end{gather}
It is  interesting to note that  \rf{fe_32}  matches the small-$t$ asymptotics of the heat kernels on the cone of $H^2$  found  in ~\cite{Bergamin:2015vxa}  when expanded   for  $k\to1$~\footnote{Here we  give   the terms proportional to $k-1$ in the expansion of eqs. (2.19)  
 (with $m^2=0,2$)  and (3.17) (with $m^2=1$) in \cite{Bergamin:2015vxa}.}
\begin{flalign}
[\tilde{K}_{\mathcal{O}_0}(t)]_{\textrm{BT}} & =\bar{K}_{-\Delta}(t)+\frac{e^{-\frac{1}{4}t}}{\sqrt{4\pi t}}\int_{0}^{\infty}dy\,e^{-\frac{y^{2}}{t}}\frac{y-\sinh y\cosh y}{\sinh^{3}y} 
\te  =-\frac{1}{2t}+O(t)\,, \\
[\tilde{K}_{\mathcal{O}_1}(t)]_{\textrm{BT}} & =\bar{K}_{-\Delta+2}(t)+\frac{e^{-\frac{9}{4}t}}{\sqrt{4\pi t}}\int_{0}^{\infty}dy\,e^{-\frac{y^{2}}{t}}\frac{y-\sinh y\cosh y}{\sinh^{3}y}
\te  =-\frac{1}{2t}+1+O(t)\,,\\
[\tilde{K}_{\mathcal{O}^2_F}(t)]_{\textrm{BT}} & =\bar{K}_{-\slashed{\nabla}^{2}+1}(t)-\frac{4e^{-t}}{\sqrt{4\pi t}}\int_{0}^{\infty}dy\,e^{-\frac{y^{2}}{t}}\frac{y\cosh y-\sinh y}{\sinh^{3}y}
\te  =-\frac{1}{t}+\frac{1}{2}+O\left(t\right) \ . 
\end{flalign}
Here we used the expansions \eqref{hk_Laplace_Seeley}-\eqref{hk_Dirac_Seeley} and performed the change of variable $y\to \sqrt{t}\, y$. 

{The non-vanishing $t^0$ term  in  the ${O}(k-1)$   correction to the total heat kernel  
 $\tilde{K}_{\rm tot}$  in \rf{fe_33}   implies the  presence of   $k$-dependent 
 logarithmic UV divergence   
 in the logarithm of the one-loop string partition function
 (implicit  also  in  \cite{Bergamin:2015vxa}). 
 The presence of this $k$-dependent UV divergence    appears  to be 
in contradiction with the fact  that the Euler number 
of  the   cone of  $AdS_2$    is the same  as  of the   disc  ($\chi=1$)  for any $k$ 
which suggests that  the UV divergence should  actually cancel  in the ratio of $k\not=1$ and $k=1$ partition functions
(as  in the latitude example of Section \ref{sec:latitude}). 
In general,  it is known that  conical singularities produce  extra  
 contributions to  the  heat coefficient 
 $a_2$   \cite{Vassilevich:2003xt,Fursaev} (cf.   \rf{ab}). 
 What happens is that   the  regular $k$-dependent bulk contribution to the   Euler  number   is cancelled against 
 the  $k$-dependent  tip   contribution.
 \foot{\label{foot_euler} The Euler number   is given   by the sum of the volume   and  boundary contributions, 
$\chi=\chi_v + \chi_b$. 
The  volume part of the Euler number $\chi_v=\frac{1}{4\pi}\int d^2 x  \sqrt g R$ 
 contains the ``regular" and ``singular" (tip) contributions: $ \chi_v=\chi_{v}^{\rm reg}+\chi_{v}^{\rm tip}$. 
 The regular part of the curvature of the metric \eqref{kcircle_metric}
 is $R_{\rm reg}=- {2 }\Omega^{-2} \partial^2_\sigma  \log \Omega =-2$,   while the tip part  is 
$R_{\rm tip}=4\,\pi\,(1-k)\, \delta^{(2)} (x)$, so that 
$\chi_{v}^{\rm reg}= \frac{1}{4\pi} \int_0^{2\pi} d\tau \int_{k^{-1}\textrm{arctanh}\,\epsilon }^\infty   d\sigma\,\Omega^2\,(\sigma)\, R_{\rm reg}
= -\frac{k}{\epsilon}  + k$  and  $\chi_{v}^{\rm  tip}=1-k$.
Thus   
$ \chi_v= 1-\frac{k}{\epsilon}$. 
The geodesic curvature  of the boundary at $\bar\sigma=k^{-1}\textrm{arctanh}\,\epsilon$
is   $
\kappa_g=\partial_\sigma\Omega^{-1}(\sigma )|_{\sigma=\bar\sigma}=\frac{1}{\epsilon}\,,
$  so that 
the boundary part of the Euler number is  
$
\chi_b=\frac{1}{2\pi}\int \,ds\,\kappa_g\equiv \frac{1}{2\pi}\int_0^{2\pi}d\tau \,\Omega(\bar\sigma)\,\kappa_g=\frac{k}{\epsilon}\,.
$
As a result, the total 
  Euler character   $\chi=\chi_v+\chi_b=1$ is finite and does not depend on $k$,  i.e.  is 
   the same as of a disc.    }

 One may then suspect that 
our perturbative approach to computation of heat kernels  may  be missing   some 
  subtleties of the heat asymptotics around the tip of 
  the cone.\footnote{See, for example~\cite{Bordag:1996fw} and references therein. We thank D. Seminara for a discussion on this point.
  Note also that our test for the scalar Laplacian in Appendix \ref{app:pert_Seeley_Laplacian} applied only to smooth manifolds.}
  Namely,    it may be  missing  the singular  tip  of the  cone   contribution to $a_2$  
so that instead of being proportional to the   full ($k$-independent)  Euler  number equal to 1 
it  appears to be  given just  by   the regular 
  bulk contribution $\chi_v^{\rm reg}=k$  (we drop  the ${1\ov \epsilon}$  part  in $\chi_v^{\textrm{reg}}$ in footnote \ref{foot_euler}
  as our usual  IR  regularization prescription).
   Explicitly, one  may then interpret  \eqref{fe_33} as the ${O}(k-1)$  term in the  total heat kernel 
    where the $t^0$ term is given by $\chi_v^{\rm reg}$, i.e. 
\be
\begin{split}\la{1337} 
\textstyle K^{{\rm reg}}_{\rm tot}(t)\te =\frac{0}{t}+\frac{k}{2}\,t^0+O(t)=\frac{0}{t}+\Big[\,\frac{1}{2}+\frac{1}{2}(k-1)\,\Big]\,t^0+O(t)~,
\end{split}
\ee
Then the effect of   proper  accounting for  the tip contribution  should be, in  particular, 
   the replacement of the $\frac{k}{2}\,t^0$  term in \rf{1337} by $\frac{1}{2}\,t^0$  and thus 
   the  cancellation of   the $1\ov 2$ term in \rf{fe_33}.  

This    suggests that the presence of the extra $\gamma$  term in \rf{3.106} (which  represents the difference with the gauge theory result) 
 may be an artifact of 
the superficial presence of   $k$-dependent  UV divergences  before the tip  contribution is taken into account. 
We leave a  careful resolution of this issue  for the future.



\iffa 
\emph{\EV{we did not say why physically the k-circle heat kernel behaves differently from those of latitude and cusp. the ultimate reason is that the heat kernel is constructed with the variations of the operators and well-chosen world-sheet coordinates, so it knows of the conical singularity. helpful comparison with the derivation of B.T. is not possible because they used exact conical geometry to construct their heat kernels, which do not reduce to ours at $k\sim 1$.}}
\emph{\EV{One observation that we find interesting is the fact that $\gamma$ is a value taken on by the digamma function $\psi(x)$, and hence it 
possess an integral representation 
\be\label{gamma}
-\gamma = \psi(1) = \int_0^{\infty} \left(\frac{e^{-t}}{t}-\frac{e^{-t}}{1-e^{-t}} \right) dt
\ee
that mimics the derivative of (fictitious) zeta-function evaluated at $s=0$. The spurious contribution proportional to $\gamma$ would be removed if were able to subtract \emph{ad hoc} \eqref{gamma} to the scalar and spinor zeta-functions \eqref{3.105}. Another possibility is speculate $\gamma=\frac{d}{ds}\left(\frac{1}{s\, \Gamma(s)} \right)_{s=0}= \frac{d}{ds}\left(\frac{1}{\Gamma(s)} \int_0^1 dt\, t^{s-1} \right)_{s=0}$. 
For introduction and here: one should study the circle with $k\sim n\in\mathbb{N}$ with our method to collect more data and refine the statements so far.}}
\fi 

\section*{Acknowledgements}

We are grateful to Amit Dekel, Valentina Giangreco Marotta Puletti, Luca Griguolo, Yi Pang, Domenico Seminara and Diego Trancanelli for useful discussions. The  work of VF is funded by DFG via the Emmy Noether Programme ``Gauge  Fields  from  Strings" 31408816. The work of AAT is supported by the ERC Advanced grant no. 290456, the  STFC Consolidated grant ST/L00044X/1 and  the Russian Science Foundation grant 14-42-00047
at Lebedev Institute. The work of  EV is funded by the FAPESP grant 2014/18634-9 and 2016/09266-1.

\appendix

\section{Perturbation theory for heat kernel and Seeley coefficients}
\label{app:pert_theory}

In this Appendix we collect some 
details on the derivation of \rf{2.7},\rf{2.8}  and show, as non-trivial consistency check, that  our perturbative expansion  reproduces
the  standard perturbation theory applied  directly 
to the Seeley coefficients of  the scalar Laplacian  operator. 

\subsection{Perturbation theory for heat kernel}
\label{app:heat_kernel_perturb}

To obtain the first correction $\tilde{K}(x,x';t)$ to the heat kernel in \eqref{expansion}, we solve equation \eqref{orderalpha} 
using  the standard method of variation of constants. We  start with the ansatz
\begin{flalign}\label{a1}
\tilde{K}_{\mathcal{O}}(x,x';t)& \equiv-\frac{\tilde{g}(x)}{2\bar{g}^{3/2}(x)}\delta^{(d)}(x-x')\mathbb{I}+\int d^d x''\sqrt{\bar{g}(x'')}\bar{K}_{\mathcal{O}}(x,x'';t)C_{\mathcal{O}}(x'',x';t)\,,\\\no
&\lim_{t\to0^+} C_{\mathcal{O}}(x,x';t)=0\,,
\end{flalign}
which guarantees that the initial condition in \eqref{orderalpha} is satisfied, and solve for $C_{\mathcal{O}}(x'',x';t)$  
\be\label{heat_equation_order_1(2)}
\!\!\!\!
\int d^d x''\sqrt{g(x'')}\bar{K}_{\mathcal{O}}(x,x'';t)~\partial_{t}C_{\mathcal{O}}(x'',x';t)=\bar{\mathcal{O}}_{x}\Big(\frac{\tilde{g}(x)}{\bar{g}^{3/2}(x)}\delta^{(d)}(x-x')\Big)-\tilde{\mathcal{O}}_{x}\bar{K}_{\mathcal{O}}(x,x';t)\,.
\ee
We now multiply  both sides by~$\sqrt{\bar{g}(x)}\bar{K}_{\mathcal{O}}(x{'''},x;-t)$~\footnote{This step is not fully rigorous because the identity \eqref{composition_law_Kbar} holds only for positive values of the proper times. In fact, the inverse heat kernel $\bar{K}^{-1}_{\mathcal{O}}(x{'''},x;t)=\bar{K}_{\mathcal{O}}(x{'''},x;-t)$ is not guaranteed to be a well-defined operator when it acts on arbitrary functions $f(x)$ taking values in a vector bundle. However, this potentially problematic operator will not enter the final formula \eqref{2.7}, which indeed contains heat kernels with only positive arguments $t'$ and $t-t'$. We could alternatively start with \eqref{2.7} and check that it is a solution of \eqref{orderzero} without the need of inverting heat kernels. A similar discussion is found in Chapter 14 of~\cite{Mukhanov}.} 
and integrate over $x$, 
using  the composition law
\be \label{composition_law_Kbar}
\int d^d x'\sqrt{\bar{g}(x')}\bar{K}_{\mathcal{O}}(x,x';t)\bar{K}_{\mathcal{O}}(x',x'';t')=\bar{K}_{\mathcal{O}}(x,x'';t+t')\,,
\qquad
t,\,t'>0\,.
\ee
With the initial condition in \eqref{orderzero}, we then obtain
\begin{eqnarray}\no 
\partial_{t}C_{\mathcal{O}}(x{'''},x';t) & = & \int d^d x\sqrt{\bar{g}(x)}\bar{K}_{\mathcal{O}}(x{'''},x;-t)\bar{\mathcal{O}}_{x}\left(\frac{\tilde{g}(x)}{\bar{g}^{3/2}(x)}\delta^{(d)}(x-x')\right)\\
 &  &- \int d^d x\sqrt{\bar{g}(x)}\bar{K}_{\mathcal{O}}(x{'''},x;-t)\tilde{\mathcal{O}}_{x}\bar{K}_{\mathcal{O}}(x,x';t)\,.  \label{C_order_1}
\end{eqnarray}
 Relabeling $x{'''}\to {x''},\,x\to x{'''},\, t\to t'$ it is straightforward to integrate over the proper time to  get
\begin{eqnarray}
C_{\mathcal{O}}(x'',x';t) & =&\int_{0}^{t}dt'\int d^dx{'''}\,\sqrt{\bar{g}(x{'''})}\bar{K}_{\mathcal{O}}(x'',x{'''};-t')\bar{\mathcal{O}}_{x{'''}}\left(\frac{\tilde{g}(x{'''})}{2\bar{g}^{3/2}(x{'''})}\delta^{(d)}(x'-x{'''})\right)\no  \\
 & & - \int_{0}^{t}dt'\int{d^dx{'''}}\sqrt{\bar{g}(x{'''})}\bar{K}_{\mathcal{O}}(x'',x{'''};-t')\tilde{\mathcal{O}}_{x{'''}}\bar{K}_{\mathcal{O}}(x{'''},x';t')\ .
\end{eqnarray}
  Substituting in \eqref{a1} this leads to the explicit integral form \eqref{2.7}, with a few more steps that employ \eqref{orderzero} and \eqref{composition_law_Kbar}. 

Next, the order $\a$  correction $\tilde{K}_{\mathcal{O}}(t)$   in \eqref{2.80} receives contributions\footnote{\label{footnote_range}
There is no correction due the integration over the $x^i$ because they range in a subset of $\mathbb{R}^n$ that is the same for $\mathcal{M}$ and $\bar{\mathcal{M}}$. Although one may argue that the expansion \eqref{expansion} needs to assume that the range of coordinates should not depend on $\alpha$, the analysis in Section \ref{sec:generalized_cusp} shows that one may allow their domain to change infinitesimally when expanding in small $\alpha$. This weaker condition on the choice of coordinates should be valid as long as the change in the integration domain in the final formula $\eqref{2.8}$ produces only small additional terms, proportional to positive powers of $\alpha$, that are eventually neglected in \eqref{2.80} at linear order in $\alpha$.} 
from both the $\alpha$-correction to volume factor $\sqrt{g(x)}$ (cf. \eqref{expansion})
 and from the $\alpha$-correction to  the heat kernel in \eqref{2.7}, i.e. 
\begin{gather}
\tilde{K}_{\mathcal{O}}(t)  =\int d^d x\frac{\tilde{g}(x)}{2\sqrt{\bar{g}(x)}}\,\tr \bar{K}_{\mathcal{O}}(x,x;t)+\int d^d x\sqrt{\bar{g}(x)}\,\tr \tilde{K}_{\mathcal{O}}(x,x;t)\,.
\end{gather}
Plugging here  the diagonal element $x=x'$ of  \eqref{2.7}, we get
\begin{eqnarray}\label{a7}
&&\tilde{K}_{\mathcal{O}}(t) =\int d^d x\frac{\tilde{g}(x)}{2\sqrt{\bar{g}(x)}}\,\tr \bar{K}_{\mathcal{O}}(x,x;t)
-\delta^{(d)}(0)\int d^d x\frac{\tilde{g}(x)}{2\bar{g}(x)}\,\tr \mathbb{I}\\
&&-\int_{0}^{t}dt' \int d^d x\sqrt{\bar{g}(x)}\int d^d x''\sqrt{\bar{g}(x'')}\,\tr \Big[\bar{K}_{\mathcal{O}}(x,x'';t-t')\tilde{\mathcal{O}}_{x''}\bar{K}_{\mathcal{O}}(x'',x;t')\Big]\no\\
\no
 &&+\int_{0}^{t}dt'\int d^d x\sqrt{\bar{g}(x)}\int d^d x''\sqrt{\bar{g}(x'')}\,\tr \Big[\bar{K}_{\mathcal{O}}(x,x'';t-t')\bar{\mathcal{O}}_{x''}\left(\frac{\tilde{g}(x'')}{2\bar{g}(x'')^{3/2}}\delta^{(d)}(x-x'')\right)\Big]\,.
\end{eqnarray}
This expression is potentially affected by two types of divergences. The first one is the $\delta^{(d)}(0)$, short-distance divergence originating from \eqref{2.7};
it  will  eventually cancel against the delta-function  in the last  integrand.
 The second   is  a possible 
 infrared  divergence   that may appear if  $\mathcal{M}$ and $\mathcal{\bar{M}}$ are non-compact. 
 In the applications to string theory in Section~\ref{sec:applications}
 the 
  volume divergences will be  regulated  by a  cutoff 
   and then subtracted through the
    renormalization prescription suggested in similar calculations in  \cite{Buchbinder:2014nia,Bergamin:2015vxa}.

To bring the integral \rf{a7}  into a more convenient form, we
shall 
 assume that $\bar{\mathcal{O}}$ is a self-adjoint operator on a vector bundle of the manifold $\bar{\mathcal{M}}$~\footnote{A natural  inner product is defined  as 
$
(f,h) \equiv \int d^d x \sqrt{g(x)} \, f^{\dagger}(x) h(x)\,.$}
\be
\int d^d x \sqrt{\bar{g}(x)} \, f^{\dagger}(x) \bar{\mathcal{O}}_x  h(x)
=
\int d^d x \sqrt{\bar{g}(x)} \,  (\bar{\mathcal{O}}_x f(x))^{\dagger}  h(x)
\,.
\ee
Combining this with \eqref{orderzero} and  setting  $t{''}=t-t'$, we rewrite the last term in \eqref{a7} as
\ba
&&
-\!\int_{0}^{t}\!\!dt^{''}\!\!\!\int d^d x\frac{\tilde{g}(x)}{2\sqrt{\bar{g}(x)}}\,\tr \big(\partial_{t^{''}}\bar{K}_{\mathcal{O}}(x,x;t^{''})\big)
\no \\
&& \ \ \ \ =
-\!\!\int d^d x  \frac{\tilde{g}(x)}{2\sqrt{\bar{g}(x)}}\, \tr \bar{K}_{\mathcal{O}}(x,x;t)
+\delta^{(d)}(0)\!\! \int d^d x  \frac{\tilde{g}(x)}{2\sqrt{\bar{g}(x)}} \tr \mathbb{I}\,,
\label{tre_4} \ea
which simplifies  \eqref{a7} to 
\begin{gather}\label{tre(5)}
\!\!\!\!\!\!\!\!\!\!\!\tilde{K}_{\mathcal{O}}(t) =
-\int_{0}^{t}dt'\int d^d x\sqrt{\bar{g}(x)}\int d^d x''\sqrt{\bar{g}(x'')}\,\tr \Big[\bar{K}_{\mathcal{O}}(x,x'';t-t')\tilde{\mathcal{O}}_{x''}\bar{K}_{\mathcal{O}}(x'',x;t')\Big]\,.
\end{gather}
We can now use the cyclicity of the trace to  write 
\begin{equation}\label{tre(6)}
\!\!\!\!\!
\tilde{K}_{\mathcal{O}}(t) =
-\int_{0}^{t}dt'\int d^d x\sqrt{\bar{g}(x)}\int d^d x''\sqrt{\bar{g}(x'')}\,\tr
\Big[
\tilde{\mathcal{O}}_{x''} \Big(
\bar{K}_{\mathcal{O}}(x'',x;t')
\bar{K}_{\mathcal{O}}(x,x';t-t')
\Big)\Big]_{x'=x''}\,,
\end{equation}
where it is understood that the limit $x'\to x''$ is taken after $\mathcal{O}_{x''}$ has acted on the argument in round brackets. 
Making use of \eqref{composition_law_Kbar} we    then get  the compact  expressions in  \eqref{2.8}.

\subsection{Perturbative expansion of  Seeley coefficients of  scalar  Laplacian
} 
\label{app:pert_Seeley_Laplacian}

Consider the scalar Laplacian  on a compact  non-singular space $\mathcal{M}$   
 with  metric $g_{ij}$ 
\be\label{laplacian_1}
\mathcal{O}  =-\frac{1}{\sqrt{g}}\partial_{i}\left(\sqrt{g}g^{ij}\partial_{j}\right)+E \ .  
\ee
Under the standard conditions  the corresponding heat kernel   may   be expanded as \cite{Gilkey,Fursaev}
\begin{gather}\label{series_ab}
K_{\mathcal{O}}(x,x;t) \simeq \frac{1}{(4\pi)^{d/2}}\sum_{k=0}^{\infty} t^{(k-d)/2}\,b_{k/2}(x)\,, \qquad 
K_{\mathcal{O}}(t) \simeq \sum_{k=0}^{\infty}t^{(k-d)/2}\,a_{k}\,,
\end{gather}
where  $a_{k}\equiv {(4\pi)^{-d/2}} \int d^d x  \sqrt{g(x)}\,  b_{k}(x)$.
Then the UV divergences of $\log \Det \mathcal{O} $   may be expressed  in terms of the Seeley coefficients $a_k$ with $k\leq d$.
 As in   Section \ref{sec:applications} we  are  interested in the case of  $d=2$, 
 here we shall concentrate only on   the   leading $a_k$.
 For  compact manifolds without boundary, odd Seeley coefficients  $a_{2l+1}$ vanish and the first non-trivial ones 
 read \cite{Gilkey, Fursaev}\footnote{The manifolds discussed in Section \ref{sec:applications} are \emph{not} compact. 
 We regularize integrations over infinite regions  by introducing  a  cutoff, i.e. a 
  boundary at a finite distance.
   This renders the integrals defining the Seeley coefficients $a_0$ and $a_2$ finite, 
   and  suggests that  we should also  consider 
   the boundary term  
   $a_1=-\frac{1}{4}\left(\frac{1}{4\pi}\right)^{\frac{d-1}{2}}\int_{\partial\mathcal{M}}\,\sqrt{g}$ proportional to the length of the boundary. 
   However, if we assume that all  the IR divergences  are completely  subtracted, that implies 
   that 
   the renormalized value of $a_1$  is effectively zero  and we can restrict consideration to  $a_0$ and the  
    (the volume part of)  $a_2$  (cf. also \cite{Drukker:2000ep,Forini:2015mca,Vescovi:2016zzu}). 
   } 
\begin{flalign}\label{ab}
a_{0} =\frac{1}{(4\pi)^{d/2}} \int d^2 x  \sqrt{g}\,  \,,
~~~\qquad \qquad 
a_{2} =\frac{1}{(4\pi)^{d/2}} \int d^2 x  \sqrt{g}\,  \big( \frac{R}{6}-E \big)  \ . 
\end{flalign}
 Consider now two conformally equivalent metrics $g_{ij}$ and $\bar{g}_{ij}$, \ \ 
  $g_{ij}=e^{2\,\alpha\,\Omega\left(x\right)}\bar{g}_{ij}$, with $\alpha$  being a small parameter. 
  Setting $E=\bar{E}+\alpha\,\tilde{E}+O (\alpha^{2})\,$   and using  \rf{expansion}  we get 
\begin{eqnarray}\label{operator_expansion}
\mathcal{O}  &=&  \bar{\mathcal{O}}+\alpha~\tilde{\mathcal{O}}+O\left(\alpha^{2}\right)\\
 & =& \Big[ -\frac{1}{\sqrt{\bar{g}}}\partial_{i} (\sqrt{\bar{g}}\bar{g}^{ij}\partial_{j})+\bar{E}\Big]+\alpha\,\Big[-\left(d-2\right)\bar{g}^{ij}\partial_{i}\Omega\partial_{j}+\frac{2\Omega}{\sqrt{\bar{g}}}\partial_{i}\left(\sqrt{\bar{g}}\bar{g}^{ij}\partial_{j}\right)+\tilde{E}\Big]+O\left(\alpha^{2}\right)\,.\nonumber 
\end{eqnarray}
Using also the expansion for the scalar curvature\footnote{Under a conformal rescaling  of the metric,
$\bar{R} \to R = e^{-2\alpha\Omega}\Big[\bar{R}-\frac{2\alpha(d-1)}{\sqrt{\bar{g}}}\partial_{i}\left(\sqrt{\bar{g}}\bar{g}^{ij}\partial_{j}\Omega\right)-\alpha^{2}(d-1)(d-2)\bar{g}^{ij}\partial_{i}\Omega\partial_{j}\Omega\Big].$
}
\begin{flalign}
R =\bar{R}+\alpha\tilde{R}+O(\alpha^{2})
  =\bar{R}-2\alpha \Big[\Omega\bar{R}+\frac{d-1}{\sqrt{\bar{g}}}\partial_{i}\left(\sqrt{\bar{g}}\bar{g}^{ij}\partial_{j}\Omega\right)\Big]+O(\alpha^{2})\,,
\end{flalign}
one obtains  to linear order in $\alpha$  the following expansion for the relevant Seeley coefficients, 
\ba
&& \qquad \qquad  a_k=\bar{a}_k+\alpha \,\tilde{a}_k+O(\alpha^{2})\,,
\\
\label{a_bar}
\!\!\!\bar{a}_{0} & =&\frac{1}{\left(4\pi\right)^{d/2}}\int d^d x\sqrt{\bar{g}}\,,\qquad 
~~\bar{a}_{2}  =\frac{1}{\left(4\pi\right)^{d/2}}\int d^d x\sqrt{\bar{g}}\Big(\frac{1}{6}\bar{R}-\bar{E}\Big)\,,\\\label{a_tilde}
\!\!\! \tilde{a}_{0} & =&\frac{d}{\left(4\pi\right)^{d/2}}\int d^d x\sqrt{\bar{g}}\,\Omega\,,\qquad  
\tilde{a}_{2}  =\frac{1}{\left(4\pi\right)^{d/2}}\int d^d x\sqrt{\bar{g}}\,\Big(\frac{d-2}{6}\,\Omega\,\bar{R}-d\,\Omega\,\bar{E}-\tilde{E}\Big).
\end{eqnarray}
As a consistency check of the perturbative approach developed in Section \ref{sec:perturbative_hk} 
let us show that  $\tilde{a}_{0}$ and $\tilde{a}_{2}$ in \eqref{a_tilde} are reproduced from 
 the  small-$t$ expansion of the   heat kernel  trace in  (\ref{2.80}),(\ref{2.8}).
%
%
%
Using  \eqref{operator_expansion}   we get 
\begin{flalign}\label{traced_hk_for_Seeley(1)}
\tilde{K}_{\mathcal{O}}(t) 
& = t\int d^d x\left\{(d-2)\sqrt{\bar{g}(x)}\bar{g}^{ij}(x)\partial^x_i \Omega(x) \left(\partial_{j}^x \bar{K}_{\mathcal{O}}(x,x';t)\right)_{x=x'}\right.\\
& \qquad~~
\left. -2\,\Omega(x) \Big[\partial^x_{i} \left(\sqrt{\bar{g}(x)}\bar{g}^{ij}(x) \partial^x_{j}\bar{K}_{\mathcal{O}}(x,x';t) \right)\Big]_{x=x'} -\sqrt{\bar{g}(x)}\tilde{E}(x)\bar{K}_{\mathcal{O}}(x,x;t) \right\}\,.\no
\end{flalign}
Integrating by parts in  the first term 
using  that the unperturbed Laplacian satisfies \eqref{orderzero}   gives 
\begin{flalign}\label{traced_hk_for_Seeley(3)}
\tilde{K}_{\mathcal{O}}(t) & = 
t \int d^d x
\Big[-(d-2)\partial_j^x \left(\sqrt{\bar{g}(x)}\bar{g}^{ij}(x) \partial_i^x \Omega(x) \right)\bar{K}_{\mathcal{O}}(x,x;t)\\
& \qquad~~~
-2\sqrt{\bar{g}(x)}\,\Omega(x)\, \partial_{t}\bar{K}_{\mathcal{O}}(x,x;t)
-\sqrt{\bar{g}(x)}\left(2\Omega(x) \bar{E}(x)+\tilde{E}(x)\right)\bar{K}_{\mathcal{O}}(x,x;t)
\Big]\,.\no
\end{flalign}
Expanding in  $t\to 0^{+}$  and using \eqref{series_ab}   we get 
\begin{flalign}\label{traced_hk_for_Seeley(4)}
{K}_{\mathcal{O}}(t) & =\bar{K}_{\mathcal{O}}(t)+\frac{\alpha}{(4\pi)^{d/2}}\Big[d\,t^{-d/2}\int d^d x\sqrt{\bar{g}(x)}\,\Omega(x)\\
& +t^{(2-d)/2}\int d^d x\sqrt{\bar{g}(x)} \te  \Big (\frac{d-2}{6}\Omega(x)\bar{R}(x)
 -d\,\Omega(x)\bar{E}(x)-\tilde{E}(x)\Big)+O(t^{(3-d)/2})\Big]+O(\alpha^2)\,. \no
\end{flalign}
Reading off the  values of  the first corrections  $\tilde{a}_0,\tilde{a}_2$ 
one  finds that they match the ones in~\eqref{a_tilde}.

\section{Heat kernels and zeta-functions  
for operators on  $H^{2}$}
\label{app:heat_kernel_zeta_function}

In this Appendix   we will review   the  known expressions  for heat kernels 
of Laplace and Dirac operators on 
the  Euclidean $AdS_2$ or  2d hyberbolic space $H^2$   with the metric
\begin{gather}\label{polar_metric}
ds^{2}=
d\rho^{2}+\sinh^{2}\rho\, d\tau^{2}\,,\qquad\rho>0\,,\qquad\tau\in\left[0,2\pi\right)\,,
\end{gather}
where  
$\tau$   parametrizes  the  $S^1$  boundary at $\rho=\infty$. 
The geodesic distance $d(x,x')$ between two points $x=(\rho,\tau)$ and $x'= (\rho',\tau')$ is   
\begin{gather}\label{polar_distance}
\cosh d(x,x')=\cosh\rho\cosh\rho'-\sinh\rho\sinh\rho'\cos(\tau-\tau')\,.
\end{gather}
\iffa 
The space is maximally symmetric and possesses three Killing vectors $\xi^i \partial_i$
\begin{gather}\label{polar_Killing}
\xi^{i}_{(1)}=(\cos \tau,\, -\coth\rho \sin \tau)\,,
\qquad
\xi^{i}_{(2)}=(\sin \tau,\, \coth\rho \cos \tau)\,,
\qquad
\xi^{i}_{(3)}=(0,\,1)
\end{gather}
that satisfy the algebra of the isometry group $SO(1,2)$ of the space.
\fi 
We will 
also considered the ``infinite-strip'' parametrization $x=\left(r,w\right)$ of   $AdS_2$    that we call $\hat H^2$, which
 has the real line  instead of $S^1$ as  its boundary
\begin{gather}\label{strip_metric}
ds^{2}=\frac{1}{\cos^{2}r}( dr^{2}+dw^{2})\,,\qquad \te  r\in\left(-\frac{\pi}{2},\frac{\pi}{2}\right)\,,\qquad w\in\mathbb{R}\ , 
\end{gather}
with   geodesic distance function
\begin{gather}\label{strip_distance}
\cosh d(x,x')=-\tan r\tan r'+\frac{\cosh(w-w')}{\cos r\cos r'}\ . 
\end{gather}
\iffa 
and Killing vectors
\begin{gather}\label{strip_Killing}
\xi^{i}_{(1)}=(\cos r \sinh w,\, -\sin r \cosh w)\,,
~
\xi^{i}_{(2)}=(0,\, 1)\,,
~
\xi^{i}_{(3)}=(\cos r \cosh w,\, -\sin r \sinh w)\,,
\end{gather}
which can be derived from \eqref{polar_metric}, \eqref{polar_distance} and \eqref{polar_Killing} through 
\fi
The change of coordinates  between the two    systems  is 
\ba\label{strip_to_polar}
&&\te \cosh\rho = \frac{\cosh w}{\cos r}\,,
\qquad
\sin\tau = \frac{\sin r}{\sqrt{\sin^{2}r+\sinh^{2} w}}\,,
\qquad
\cos\tau = \frac{\sinh w}{\sqrt{\sin^{2}r+\sinh^{2} w}}\ , \ \ \ 
\\  &&\te \label{polar_to_strip}
\tan r  =\sinh\rho \sin\tau\,, \qquad
\tanh w  =\tanh\rho \cos\tau\,.
\ea
We shall consider a 
Laplace type   operator  acting on function in a vector bundle 
$-\frac{1}{\sqrt{g}}(\partial_{i}+A_{i})\left(\sqrt{g}g^{ij}(\partial_{j}+A_{j})\right)+E 
$
and also a  Dirac type acting on two-dimensional spinors
$ 
- i\slashed{\nabla} +V \equiv i e^{~i}_a\Gamma^{a}\nabla_{i}+V ,   $
where the  spinor derivative  is $\nabla_i\equiv\partial_{i}+\frac{1}{4}\omega_{\phantom{ab}c}^{ab}e^c_{~i}\Gamma_{ab}$.\foot{
The  coordinate indices   are $i,j,...=1,2$,  the  indices of  the local orthonormal frame  are  $a,b,...=1,2$ 
and $\alpha,\beta,...=1,2$  are  the indices of  the spinor bundle over $\mathcal{M}$
(we follow  mainly   the conventions  of  Appendix A of~\cite{Forini:2015bgo}).}
  $e^a_{~i}$  
  is  the zweibein,   
   $\omega_{\phantom{ab}c}^{ab}$ is the spin connection and $\Gamma_{a}$ are hermitian $SO(2)$ Dirac matrices
\begin{gather}\label{pauli_1_2}
\Gamma_{1}=\sigma_{1}\,,\qquad \Gamma_{2}=\sigma_{2}\,,\qquad\Gamma_{3}=-i\,\Gamma^1 \Gamma^2 =\sigma_{3}\,,\qquad\{\Gamma_a\,, \Gamma_b\}=2\delta_{ab} \mathbb{I}_2\,,
\end{gather}
The  explicit expressions for the  scalar Laplacian in the two coordinates \eqref{polar_metric} and \eqref{strip_metric}  are 
\ba 
&&\Delta \equiv \frac{1}{\sqrt{{g}}}\partial_{i}\left(\sqrt{{g}}{g}^{ij}\partial_{j}\right) \ , \qquad \qquad
\label{polar_Laplace_operator}
\Delta_{\rho,\tau} = \partial_{\rho}^{2}+{\coth\rho}\, \partial_{\rho}+{\sinh^{-2}\rho}\, \partial_{\tau}^{2}\,,
\\
\label{strip_Laplace_operator}
&&\qquad\qquad\qquad\qquad \qquad \Delta_{r,w} = \cos^{2}r\left(\partial_{r}^{2}+\partial_{w}^{2}\right)\,. 
\ea
The operator $-\Delta$ is hermitian with a continuous spectrum of positive eigenvalues~$\lambda\in (\frac{1}{4},\,\infty]$. 
The corresponding heat kernel for  the massive operator $-\Delta + m^2$ is ~\cite{Camporesi:1990wm, Camporesi:1994ga, Chavel, JonesKucerovsky, Jones} 
\be\label{B.11}
{K}_{-\Delta + m^2 }(x,x';t)  ={\te \frac{1}{2\pi}}\int_{0}^{\infty}dv\,v\tanh(\pi v)P_{-\frac{1}{2}+iv}(\cosh d(x,x')) \ e^{-t(v^{2}+\frac{1}{4} + m^2)}\,,
\ee
where 
the Legendre function   is indexed by $v\equiv\sqrt{ \lambda-\frac{1}{4}}>0$ 
and the geodesic distance is given by 
\eqref{polar_distance}   and   \eqref{strip_distance} in the coordinate sets \eqref{polar_metric} and \eqref{strip_metric} respectively. 

The Dirac operator $-i \slashed{\nabla}$ 
has the following explicit form
\begin{flalign}\label{polar_Dirac_operator}
-i\slashed{\nabla}_{\rho,\tau} &=-i\Gamma^{1}\left(\partial_{\rho}+\ha {\coth\rho} \right)-{i}{\sinh^{-1} \rho}\ \Gamma^{2}\partial_{\tau}\,,\\
\label{strip_Dirac_operator}
-i\slashed{\nabla}_{r,w} &= -i\Gamma^{1}\left(\cos r\partial_{r}+\ha{\sin r}\right)-i\cos r\,\Gamma^{2}\partial_{w} \,
\end{flalign}
in the two coordinate sets \eqref{polar_metric}  and \eqref{strip_metric}. 
The spinor heat kernel 
for the Dirac operator  with  a constant chiral mass term 
 $-i \slashed{\nabla}_{\rho,\tau}+m\, \Gamma^3$  (with $m\,\in \mathbb{R}$)
that  satisfies the heat equation for $-\slashed{\nabla}^2 +m^2$ 
can be written in a coordinate-independent form as the product of the parallel spinor propagator $U(x,x')$ and a scalar function of the geodesic distance $d(x,x')$ between the two points $x,x'$ \cite{Camporesi:1995fb}
\begin{flalign}\label{b13}
{K}_{-\slashed{\nabla}^{2} +m^2}(x,x';t) & ={\te \frac{1}{2\pi} }\, U(x,x')\int_{0}^{\infty}dv\:v\ \textstyle{\coth\pi v\,\cosh \Big(\frac{1}{2}{d(x,x')}\Big)}\\
&\qquad\qquad\qquad \times {}_{2}F_{1}\Big(\textstyle{iv+1,\,-iv+1,\,1,\,\frac{1}{2}-\frac{1}{2}\cosh ( d(x,x'))}\Big)e^{-t \,(v^{2} + m^2) }\,.\no
\end{flalign}
The unitary $2\times 2$ matrix $U(x,x')$ 
is the regular solution of the parallel transport equation~\cite{Camporesi:1992tm}
\begin{gather}
n^{i}(u)\nabla_{i}U(x(0),x(u))=0,\,
\qquad \qquad
U\left(x(0),x(0)\right)=\mathbb{I}_{2}\,,
\end{gather}
where   $n_i(u)= \partial_{i}d(x(0),x(u))$  is the unit vector  tangent to the shortest geodesic $x(u)$  between 
 $x(0)=x'$ and $x(1)=x$.  
 The explicit expression of $U(\rho,\tau,\rho',\tau')$ for the metric \eqref{polar_metric}
 in the matrix representation \eqref{pauli_1_2}  reads ~\cite{Bergamin:2015vxa}
\begin{eqnarray}\label{U_polar_1}
U(\rho,\tau,\rho',\tau') &=& \mathbb{I}_{2}\cos\theta(\rho,\tau,\rho',\tau')+i\Gamma_{3}\sin\theta(\rho,\tau,\rho',\tau')\,,\\\label{U_polar_2}
\theta(\rho,\tau,\rho',\tau')&\equiv&
\textrm{arctan}\textstyle{(\frac{\cosh(\frac{\rho+\rho'}{2})\tan(\frac{\tau-\tau'}{2})}{\cosh(\frac{\rho-\rho'}{2})})}\,.
\end{eqnarray}
\iffa 
The effect of switching on a constant chiral mass term in the Dirac operator $-i \slashed{\nabla}_{\rho,\tau}+m\, \Gamma^3$ with $m\,\in \mathbb{R}$ is to insert $e^{-t m^2}$ in \eqref{b13}
\begin{flalign}\label{hk_polar_Dirac_massive}
{K}_{-\slashed{\nabla}^{2}+m^2}(\rho,\tau,\rho',\tau';t) & =\frac{U(\rho,\tau,\rho',\tau')}{2\pi}\int_{0}^{\infty}dv\:v\, \coth\pi v\,\cosh \textstyle{\Big(\frac{1}{2}{d(\rho,\tau,\rho',\tau')}\Big)}\\
&\qquad \times {}_{2}F_{1}\big(\textstyle{iv+1,\,-iv+1,\,1,\,\frac{1}{2}-\frac{1}{2}\cosh ( d(\rho,\tau,\rho',\tau'))}\big)e^{-t (v^{2}+m^2)}\,, \no
\end{flalign}
with the geodesic distance in \eqref{polar_distance}. 
\fi 
The expression  in \rf{b13}  is the  solution of the heat equation  
\be\label{b133}
\begin{split}
(\partial_t-\slashed{\nabla}_{\rho,\tau}^2+m^2){K}_{-\slashed{\nabla}^2+m^2}(\rho,\tau,\rho',\tau';t)=0\,,~~ \\
\lim_{t\to 0^+} {K}_{-\slashed{\nabla}^2+m^2}(\rho,\tau,\rho',\tau';t) ={ \frac{\delta(\rho-\rho')\delta(\tau-\tau')}{\sinh\rho}} \, \mathbb{I}_2\,.
\end{split}
\ee
To change the coordinates to the  infinite-strip parametrization \eqref{strip_metric} through \eqref{polar_to_strip} 
 recall that spinors and the parallel spinor propagator are scalars under the diffeomorphisms, 
while under local rotations of the orthonormal frame they transform as
\be\label{LLT}
\begin{split}
\psi(x)
 \to
\psi(\hat{x})=S(\hat{x}) \psi(x(\hat{x}))\,,
\qquad
U(x,x')
\to
U(\hat{x},\hat x') =
S(\hat{x}) U(x(\hat{x}),x'(\hat x')) S^\dagger(\hat x')\,,
\\
\qquad
S(\hat{x})\,\Gamma^{\hat{a}}\,S^{\dagger}(\hat{x})=\Lambda_{\phantom{a}b}^{\hat{a}}(\hat{x})\,\Gamma^{b}\,,
\qquad
 S(\hat{x})\in Spin(2)\,,
\qquad \Lambda(\hat{x}) \in SO(2)\,.
\end{split}
\ee
Here  $x=(\rho,\tau)$ and $\hat{x}=(r,w)$  represent one  point 
 and  $x'=(\rho',\tau')$ and $\hat x'=(r',w')$   another  point. 
The tangent frame rotation $\Lambda(\hat{x})$, satisfying
$
e^{\hat{a}}_{~\hat{i}}(\hat{x})=
 \Lambda^{\hat{a}}_{~b}(\hat{x}) \,
 \frac{\partial x^j(\hat{x})}{\partial \hat{x}^{\hat{i}}} \,
 e^{b}_{~j}(x(\hat{x}))
$, 
reads
\be \!\!\!
\Lambda_{\phantom{a}b}^{\hat{a}}(\hat{x}) \equiv
\Big(\begin{array}{cc}
\cos\delta(\hat{x}) & \sin\delta(\hat{x})\\
-\sin\delta(\hat{x}) & \cos\delta(\hat{x})
\end{array}\Big)\,,~\sin\delta(\hat{x}) = \textstyle\frac{\cos r\sinh w}{\sqrt{\sin^{2}r+\sinh^{2}w}} \,,
~
\cos\delta(\hat{x}) = \textstyle\frac{\sin r\cosh w}{\sqrt{\sin^{2}r+\sinh^{2}w}}\,,
\ee
and the associated unitary rotation on the spinor indices is
\begin{gather}
S(\hat{x})\label{S}
=\textstyle{\cos\big(\frac{\delta(\hat{x})}{2}\big) \mathbb{I}_{2}+i\Gamma^{3}\sin\big(\frac{\delta(\hat{x})}{2}\big)}\,.
\end{gather}
The parallel spinor  propagator in the infinite-strip coordinates \eqref{strip_metric} is thus explicitly
\be\label{U_strip}
\begin{split}
U(r,w,r',w')  
&=\textstyle{\mathbb{I}_{2}\cos\big(\theta(\rho,\tau,\rho',\tau')+\frac{1}{2}\delta(r,w)-\frac{1}{2}\delta(r',w')\big)}\\
&\quad +\textstyle{i\,\Gamma_{3}\sin\big(\theta(\rho,\tau,\rho',\tau')+\frac{1}{2}\delta(r,w)-\frac{1}{2}\delta(r',w')\big)\,.}
\end{split}
\ee
In these coordinates the spinor heat kernel ${K}_{-\slashed{\nabla}^{2}+m^2}(r,w,r',w';t)$ is given by \rf{b13}  with $d(x,x')$ in \rf{strip_distance} and 
$U$ in  \rf{U_strip}
\iffa 
becomes
\begin{flalign}\label{hk_strip_Dirac_massive}
{K}_{-\slashed{\nabla}^{2}+m^2}(r,w,r',w';t) & =\frac{U(r,w,r',w')}{2\pi}\int_{0}^{\infty}dv\:v\coth\pi v\,\cosh\textstyle{ \left(\frac{1}{2}{d(r,w,r',w')}\right)}\\
&\qquad \times {}_{2}F_{1}\textstyle{\Big(iv+1,\,-iv+1,\,1,\,\frac{1}{2}-\frac{1}{2}\cosh ( d(r,w,r',w'))\Big)}e^{-t (v^{2}+m^2) }\,.\no
\end{flalign}
with the distance function in \eqref{strip_distance} and heat equation
\fi  
 satisfies 
\be\label{heat_equation_strip_dirac_massive}
\begin{split}
&\left(\partial_t-\slashed{\nabla}_{r,w}^2+m^2\right){K}_{-\slashed{\nabla}^2+m^2}(r,w,r',w';t)=0\,,\\
&\lim_{t\to 0^+} {K}_{-\slashed{\nabla}^2+m^2}(r,w,r',w';t) = \cos r \, \delta(r-r')\delta(w-w') \, \mathbb{I}_2\,. 
\end{split}
\ee

\subsection{Zeta-functions of the Laplace and Dirac operator}
\label{app:zeta_functions}

The finite parts of the determinants of the massive Laplace and Dirac operator in $H^2$ are 
given by the derivative  of the corresponding spectral zeta-function   which itself   can be expressed 
in terms of the functional trace of the  heat kernels \eqref{B.11} and \eqref{b13}
 (see also Appendix B of~\cite{Buchbinder:2014nia} and \cite{Bergamin:2015lwi}). 
The integrated heat kernel for the massive Laplace operator $-\Delta+m^2$ is~\cite{Camporesi:1990wm, Camporesi:1994ga}
\begin{flalign}\label{traced_hk_Laplace}
{K}_{-\Delta+m^{2}}\left(t\right) & =\frac{V_{H^{2}}}{2\pi}\int_{0}^{\infty}dv\,v\tanh\left(\pi v\right)e^{-t\left(v^{2}+\frac{1}{4}+m^{2}\right)}\ , 
\end{flalign}
and for the square of the massive Dirac operator $-i\slashed{\nabla}+m \Gamma^3$ is~\cite{Camporesi:1992tm,Camporesi:1995fb}
\begin{flalign}\label{traced_hk_Dirac}
{K}_{-\slashed{\nabla}^{2}+m^{2}}\left(t\right) & =\frac{V_{H^{2}}}{\pi}\int_{0}^{\infty}dv\,v\coth\left(\pi v\right)e^{-t\left(v^{2}+m^{2}\right)}\,.
\end{flalign}
The Seeley coefficients can be read off from the small-$t$ expansions
\begin{flalign}\label{hk_Laplace_Seeley}
\bar{K}_{-\Delta+m^{2}}\left(t\right) & =\frac{V_{H^{2}}}{2\pi}\left[\frac{e^{-\left(\frac{1}{4}+m^{2}\right)t}}{2t}-\int_{0}^{\infty}dv\,\frac{2v}{e^{2\pi v}+1}e^{-t\left(v^{2}+\frac{1}{4}+m^{2}\right)}\right]\\
 & =\frac{V_{H^{2}}}{4\pi}\left[\frac{1}{t}-\left(\frac{1}{3}+m^{2}\right)+O\left(t\right)\right]\nonumber \\
\label{hk_Dirac_Seeley}
\bar{K}_{-\slashed{\nabla}^{2}+m^{2}}\left(t\right) & =\frac{V_{H^{2}}}{\pi}\left[\frac{e^{-m^{2}t}}{2t}+\int_{0}^{\infty}dv\frac{2v}{e^{2\pi v}-1}e^{-t\left(v^{2}+m^{2}\right)}\right]\\
 & =\frac{V_{H^{2}}}{4\pi}\left[\frac{2}{t}+\left(\frac{1}{3}-2m^{2}\right)+O\left(t\right)\right]\nonumber
\end{flalign}
by replacing $\tanh(\pi v) =1-{2}/({e^{2\pi v}+1})$ and $\coth(\pi v) =1+{2}/({e^{2\pi v}-1})$, and they agree with the general results in \cite{Gilkey}. The zeta-function
for the massive  Laplace operator  is  
\begin{gather}
{\zeta}_{-\Delta+m^{2}}\left(s\right) =\frac{V_{H^{2}}}{2\pi}\int_{0}^{\infty}dv\frac{v\tanh\pi v}{\left(v^{2}+m^{2}+\frac{1}{4}\right)^{s}} \ . 
\end{gather}
This  expression is  valid for $\textrm{Re}\,s>1$. For the analytic continuation to a neighbourhood of  $s=0$, we  first 
use  $\tanh(\pi v) =1-{2}/({e^{2\pi v}+1})$   so that 
\begin{gather}
{\zeta}_{-\Delta+m^{2}}\left(s\right)  =\frac{V_{H^{2}}}{2\pi}\Big[
\int_{0}^{\infty}dv
\frac{v}{\left(v^{2}+m^{2}+\frac{1}{4}\right)^{s}}
-\int_{0}^{\infty}dv
\frac{2v}{\left(e^{2\pi v}+1\right)\left(v^{2}+m^{2}+\frac{1}{4}\right)^{s}}\Big]\,,
\end{gather}
where the second integral is exponentially convergent for large $v$ at $s=0$. The analytic continuation of the first integral can be easily found
 giving 
\begin{gather}
{\zeta}_{-\Delta+m^{2}}\left(s\right) =\frac{V_{H^{2}}}{2\pi}\Big[\frac{\left(m^{2}+\frac{1}{4}\right)^{1-s}}{2\left(s-1\right)}-2\int_{0}^{\infty}dv\frac{v}{\left(e^{2\pi v}+1\right)\left(v^{2}+m^{2}+\frac{1}{4}\right)^{s}}\Big]\,.
\end{gather}
Then taking the derivative with respect to $s$ and using the integral   in 
 \eqref{formula1}, we obtain
\begin{flalign}\label{ze_scalar}
{\zeta'}_{-\Delta+m^{2}}\left(0\right)  
 &=\frac{V_{H^{2}}}{2\pi}\Big[\frac{1+\log2}{12}-\log A+\int_{0}^{m^{2}+1/4}dx\,\psi\left(\sqrt{x}+\frac{1}{2}\right)\Big]  \,.
\end{flalign}
Similarly, for \eqref{traced_hk_Dirac}  using $\coth(\pi v) =1+{2}/({e^{2\pi v}-1})$ and \eqref{formula2} we  get~\footnote{Compared to~\cite{Buchbinder:2014nia},  here we do not include the minus sign of  fermionic statistics of the spinor fields in the definition of the zeta-function, but we account for it in the sum over the 
scalar and spinor contributions to the one-loop effective actions \eqref{3.9_order_0}, \eqref{3.57_order_0} and \eqref{kcircle_one_loop_eff_action_order_0}. We also recall that the spinor heat kernel in Appendix B of~\cite{Buchbinder:2014nia} and~\cite{Bergamin:2015vxa} is for Majorana fermions, so the integrated heat kernel and zeta-function include an extra factor of $1/2$ with respect to the expressions \eqref{traced_hk_Dirac} and \eqref{ze_spinor} for Dirac spinors  derived from the heat kernel in~\cite{Camporesi:1992tm,Camporesi:1995fb}.}
\be\label{ze_spinor}
\begin{split}
{\zeta}_{-\slashed{\nabla}^{2}+m^{2}}\left(s\right) & =\frac{V_{H^{2}}}{\pi}\Big[\,\frac{ (m^{2} )^{1-s}}{2\left(s-1\right)}+2\int_{0}^{\infty}dv\frac{v}{\left(e^{2\pi v}-1\right)\left(v^{2}+m^{2}\right)^{s}}\,\Big]\,, \\
{\zeta'}_{-\slashed{\nabla}^{2}+m^{2}}\left(0\right) & =\frac{V_{H^{2}}}{\pi}\Big[-\frac{1}{6}+2\log A+\sqrt{m^{2}}+\int_{0}^{m^{2}}dx\,\psi\left(\sqrt{x}\right)\Big]\,.
\end{split}
\ee
As for any homogeneous space, for which the heat kernel ${K}_{\mathcal{O}}(x,x;t)$ is independent of the point $x$, the integrated heat kernels above are all proportional to the volume of $H^2$. The latter  has to replaced by its renormalized value, as  discussed   in 
Section \ref{sec:applications}. 


\subsection{Useful integrals}
Here we 
 collect  some  integrals useful for the computation of the regularized determinants of the 
  Laplace and Dirac operators in $H^2$  (see also~\cite{Drukker:2000ep,Buchbinder:2014nia,Bergamin:2015lwi}). 
  Below,  $c$ is some  non-negative constant, $A\approx1.282$ is the Glaisher constant, $\gamma\approx0.577$
is the Euler-Mascheroni constant and $\psi\left(x\right)\equiv\frac{d}{dx}\log\Gamma\left(x\right)$
is the digamma function:
\begingroup
\allowdisplaybreaks
\begin{flalign}
&\int_{0}^{\infty}dv\frac{v\log\left(v^{2}+c\right)}{e^{2\pi v}+1} =  {\te 
{\frac{c}{4}\left(1-\log c\right)+\frac{1+\log2}{24}-\frac{\log A}{2}+ \ha  }  }
{ \int_{0}^{c}dx\,\psi(\sqrt{x}+\ha )}\label{formula1}\\
&\int_{0}^{\infty}dv\frac{v\log\left(v^{2}+\frac{1}{4}\right)}{e^{2\pi v}+1}  =\textstyle{\frac{5}{48}-\frac{\log2}{8}+\log A-\frac{\log\pi}{4}}\label{formula3}\\
&\int_{0}^{\infty}dv\frac{v\log\left(v^{2}+\frac{9}{4}\right)}{e^{2\pi v}+1}  =\textstyle{\frac{77}{48}+\frac{3\log2}{8}-\frac{9\log3}{8}+\log A-\frac{3\log\pi}{4}}\label{formula4}\\
&\int_{0}^{\infty}dv\frac{v}{\left(e^{2\pi v}+1\right)\left(v^{2}+c\right)}  =\textstyle{-\frac{\log c}{4}+\frac{1}{2}\psi\left(\sqrt{c}+\frac{1}{2}\right)}\\
&\int_{0}^{\infty}dv\frac{v}{\left(e^{2\pi v}+1\right)\left(v^{2}+\frac{1}{4}\right)}  =\textstyle{\frac{\log2}{2}-\frac{\gamma}{2}}\label{formula7}\\
&\int_{0}^{\infty}dv\frac{v}{\left(e^{2\pi v}+1\right)\left(v^{2}+\frac{9}{4}\right)}  =\textstyle{-\frac{\log\frac{3}{2}}{2}+\frac{1}{2}-\frac{\gamma}{2}}\label{formula6}\\
&\int_{0}^{\infty}dv\frac{v\log\left(v^{2}+c\right)}{e^{2\pi v}-1}  ={\te  \frac{c}{4}\left(\log c-1\right)+\frac{1}{12}-\log A-\frac{\sqrt{c}}{2}-\frac{1}{2}} \int_{0}^{c}dx\,\psi\left(\sqrt{x}\right)
\label{formula2}\\
&\int_{0}^{\infty}dv\frac{v\log\left(v^{2}+1\right)}{e^{2\pi v}-1}  =\textstyle{-\frac{2}{3}+\frac{\log2}{2}-\log A+\frac{\log\pi}{2}}\label{formula5}\\
&\int_{0}^{\infty}dv\frac{v}{\left(e^{2\pi v}-1\right)\left(v^{2}+c\right)}  =\textstyle{\frac{\log c}{4}-\frac{1}{4\sqrt{c}}-\frac{1}{2}\psi\left(\sqrt{c}\right)}\\
&\int_{0}^{\infty}dv\frac{v}{\left(e^{2\pi v}-1\right)\left(v^{2}+1\right)}  =\textstyle{-\frac{1}{4}+\frac{\gamma}{2}}\label{formula8}
\end{flalign}
\endgroup%

\iffa 
\section{Curvature and Euler number of  the cone of $AdS_2$}
\label{app:cone}

The curvature corresponding to the metric \eqref{kcircle_metric} has two contributions: a regular (bulk)  one 
and a  singular (tip) one  ($\Omega = { k \ov \sinh  k \s}; \ x= (\s, \tau)$) 
\be
 R_{\rm reg}=- {2 }\Omega^{-2} \partial^2  \log \Omega =-2\,,\qquad\qquad
R_{\rm tip}=4\,\pi\,(1-k)\, \delta^{(2)} (x) ~. 
\ee
The Euler number   is given   by the sum of the volume   and  boundary contributions, 
$\chi=\chi_v + \chi_b$. 

The  volume part of the Euler number $\chi_v=\frac{1}{4\pi}\int d^2 x  \sqrt g R$ 
thus contains the regular and singular tip contributions
\begingroup
\allowdisplaybreaks
\begin{flalign}\label{chivtot}
 \chi_v&=\chi_{v}^{\rm reg}+\chi_{v}^{\rm tip}= 1-\frac{k}{\epsilon}\ , \\\label{chivsmooth}
\chi_{v}^{\rm reg}&= \frac{1}{4\pi} \int_0^{2\pi} d\tau \int_{k^{-1}\tanh ^{-1}\epsilon }^\infty   d\sigma\,\Omega^2\,(\sigma)\, R_{\rm reg}
= k-\frac{k}{\epsilon}\ , \\\label{chivtip}\ 
\chi_{v}^{\rm  tip}&=1-k \ . 
\end{flalign}
\endgroup
Given the geodesic curvature  of the boundary at $\bar\sigma=k^{-1}\tanh ^{-1}\epsilon$
\be
\kappa_g=\partial_\sigma\Omega^{-1}(\sigma )|_{\sigma=\bar\sigma}=\frac{1}{\epsilon}\,,
\ee
the boundary part of the Euler number reads  ($ds$ is the invariant line element at the boundary)
\be
\chi_b=\frac{1}{2\pi}\int \,ds\,\kappa_g\equiv \frac{1}{2\pi}\int_0^{2\pi}d\tau \,\Omega(\bar\sigma)\,\kappa_g=\frac{k}{\epsilon}\,.
\ee
As a result, the total 
  Euler character   $\chi=\chi_v+\chi_b=1$ is finite and does not depend on $k$,  i.e.  is 
   the same as of a disc.

\fi

\bibliographystyle{nb}
\bibliography{Ref_Heat_kernel}

\end{document}